\documentclass[superscriptaddress,aps,twocolumn,longbibliography]{revtex4-1}
\usepackage{amsmath,amssymb}
\usepackage{graphicx}
\usepackage{dcolumn}
\usepackage{enumitem}
\usepackage{float}
\usepackage{bm,dsfont}
\usepackage{multirow}
\usepackage{simpler-wick}
\usepackage{cuted}
\usepackage{color}
\usepackage{hyperref}
\hypersetup{
	colorlinks = true,
	linkcolor = [rgb]{0.70,0.13,0.13},
	citecolor = [rgb]{0.13,0.55,0.13},
	urlcolor = [rgb]{0.25, 0.41, 0.88}}
\newcommand{\bra}[1]{{\langle #1|}}
\newcommand{\ket}[1]{{|#1 \rangle}}
\newcommand{\braket}[2]{{\langle #1|#2\rangle}}
\newcommand{\dd}{\mathrm{d}}
\newcommand{\ii}{\mathrm{i}}
\newcommand{\id}{\mathds{1}}
\newcommand{\U}{\mathrm{U}}

\newcommand{\dsE}{\mathbb{E}}
\newcommand{\dsZ}{\mathbb{Z}}
\newcommand{\dsR}{\mathbb{R}}

\newcommand{\dsC}{\mathbb{C}}

\newcommand{\scL}{\mathcal{L}}

\newcommand{\scO}{\mathcal{O}}

\newcommand{\Tr}{\operatorname{Tr}}

\newcommand{\sech}{\operatorname{sech}}
\newcommand{\arccosh}{\operatorname{arccosh}}
\newcommand{\OTOC}{\operatorname{OTOC}}

\newcommand{\smat}[1]{\left[\begin{smallmatrix}#1\end{smallmatrix}\right]}

\newcommand{\dia}[3]{\raisebox{#3pt}{\includegraphics[height=#2pt]{dia_#1}}}
\newcommand{\eq}[1]{\begin{equation}#1\end{equation}}
\newcommand{\eqs}[1]{\begin{equation}\begin{split}#1\end{split}\end{equation}}
\newcommand{\eqnref}[1]{Eq.\,\eqref{#1}}
\newcommand{\figref}[1]{Fig.\,\ref{#1}}
\newcommand{\tabref}[1]{Tab.\,\ref{#1}}
\newcommand{\secref}[1]{Sec.\,\ref{#1}}\newcommand{\appref}[1]{Appendix\,\ref{#1}}
\newcommand{\refcite}[1]{Ref.\,\cite{#1}}

\usepackage[mathlines]{lineno}

\begin{document}
\title{Multi-Region Entanglement in Locally Scrambled Quantum Dynamics}
\author{A. A. Akhtar}
\author{Yi-Zhuang You}
\affiliation{Department of Physics, University of California San Diego, La Jolla, CA 92093, USA}
\date{\today}
\begin{abstract}
We study the evolution of multi-region bipartite entanglement entropy under locally scrambled quantum dynamics. We show that the multi-region entanglement can significantly modify the growth of single-region entanglement, whose effect has been largely overlooked in the existing literature. We developed a novel theoretical framework, called the entanglement feature formalism, to organize all the multi-region entanglement systematically as a sign-free many-body state. We further propose a two-parameter matrix product state (MPS) ansatz to efficiently capture the exponentially many multi-region entanglement features. Using these tools, we are able to study the multi-region entanglement dynamics jointly and represent the evolution in the MPS parameter space. By comparing the dynamical constraints on the motion of entanglement cuts, we are able to identify different quantum dynamics models in a unifying entanglement feature Hamiltonian. Depending on the quantum dynamics model, we find that multi-region effects can dominate the single region entanglement growth and only vanish for Haar random circuits. We calculate the operator-averaged out-of-time-order correlator based on the entanglement feature Hamiltonian and extract the butterfly velocity from the result. We show that the previously conjectured bound between the entanglement velocity and the butterfly velocity holds true even under the influence of multi-region entanglement. These developments could enable more efficient numerical simulations and more systematic theoretical understandings of the multi-region entanglement dynamics in quantum many-body systems.
\end{abstract}
	
\pacs{Valid PACS appear here} 

\maketitle
	
\section{Introduction}\label{intro}

The entanglement dynamics in quantum many-body systems has attracted much research interest\cite{Zyczkowski2002DQE,Kim2013BSEDNS,Chandran2015Finite,Kaufman2016QTTEIMS,Ho2017Entanglement,Zhou2017Operator,Mezei2017On,Nahum2017Quantum,Jonay2018CDOSE,Vijay2018Finite-Temperature,Zhou2019Emergent,Keyserlingk2018Operator,Nahum2018Operator,Nahum2018Dynamics,Chan2018Spectral,Rakovszky2018Diffusive,Khemani2018Operator,Chan2018Solution,Mezei2018Membrane,Bertini2019ESMMMMQC,Rakovszky2019Entanglement}. For an isolated quantum many-body system described by a pure state $\ket{\Psi}$, the bipartite quantum entanglement can be quantified by the entanglement entropy $S_\Psi(A)$, which characterizes the amount of entanglement between a region (subsystem) $A$ and its complement. While much progress has been made in understanding the entanglement entropy growth \cite{Mezei2017On,Nahum2017Quantum,Keyserlingk2018Operator,Jonay2018CDOSE,Rakovszky2019Entanglement}, most work is restricted to studying single-region entanglement, namely, when $A$ is a single contiguous region. However, very little is known about \emph{multi-region} entanglement, where the entangling region $A$ can consist of several disjoint subregions. The multi-region entanglement has been used to reveal distinctive entanglement structures within volume-law states\cite{Fan2020Self-Organized,Vijay2020Measurement-Driven}. In this work, we will focus on the multi-region entanglement and analyze its effects on the entanglement dynamics. 

Consider a system with $N$ qudits, where each qudit corresponds to a $d$-dimensional Hilbert space. Then the total number of choices for the entangling region $A$ is $2^N$. To organize all the $2^N$ corresponding entanglement entropies systematically, \refcite{You2018Entanglement,Kuo2019Markovian} introduced the \emph{entanglement feature} state as a fictitious many-body  state $\ket{W_\Psi}=\sum_{A}W_\Psi(A)\ket{A}$ that stores the entanglement features $W_\Psi(A)\equiv e^{-S_\Psi(A)}$ as coefficients of the state vector. In this way, all regions (both single- and multi-regions) are treated on equal footing. The entanglement feature state $\ket{W_\Psi}$ characterizes all the bipartite entanglement of the physical state $\ket{\Psi}$. \refcite{Fan2020Self-Organized}  suggested that the entanglement feature state $\ket{W_\Psi}$ can be compressed in terms of the matrix product state (MPS) \cite{Verstraete2008Matrix,Schollwock2011DRGMPS,ORUS2014117}, which allows us to use much less parameters (polynomial in $N$) to approximately parametrize exponentially many  entanglement entropies in all $2^N$ regions. In this work, we further develop this idea and propose a two-parameter MPS ansatz for $\ket{W_\Psi}$, which can capture both the area-law and volume-law scaling for the single-region entanglement, while providing a systematic modeling for the multi-region entanglement at the same time. We interpret the physical meaning of the MPS parameters and use them to define a two-dimensional phase space for different entanglement structures. 

The entanglement feature formalism provides a unified approach to study the entanglement dynamics in a large class of models. As proven in \refcite{Kuo2019Markovian}, the time evolution of the entanglement feature state is Markovian for any locally-scrambled quantum dynamics, which allows us to predict the entanglement dynamics for all regions at once by solving an (imaginary-time) Schr\"odinger equation $-\partial_t\ket{W_\Psi}=H_\text{EF}\ket{W_\Psi}$ (or its discrete version). Many of the quantum dynamics studied in the literature are locally-scrambled, including random unitary circuits\cite{Nahum2017Quantum,Zhou2019Emergent,Keyserlingk2018Operator,Nahum2018Dynamics}, random Hamiltonian dynamics\cite{Vijay2018Finite-Temperature,You2018Entanglement}, and quantum Brownian dynamics\cite{Lashkari2013TFSC,Xu2018LQFS,Gharibyan2018ORMBSS,Zhou2019ODBQC}. They all share the locally-scrambled property that every step of the time-evolution is drawn from a random unitary ensemble which is invariant under local (on-site) basis transformations (as if the locally basis are separately scrambled in each step). Using the entanglement feature formalism, we can explore how the corresponding MPS parameter of $\ket{W_\Psi}$ evolves, as the quantum system thermalizes from an initial product state. We can also study the effect of multi-region entanglement on the entanglement dynamics. We found that the dynamics of single-region and multi-region entanglement are generally coupled together. The only known exception is the random unitary circuit, where the dynamical equation is closed within the single-region sector. For generic locally-scrambled quantum dynamics, we derived the multi-region correction to the single-region entanglement dynamics.

Within the entanglement feature formalism, we can also study the operator spreading\cite{Zhou2017Operator,Mezei2017On,Jonay2018CDOSE,Keyserlingk2018Operator,Nahum2018Operator,Khemani2018Operator,Nie2018Signature,Parker2018UOGH,Gopalakrishnan2018Hydrodynamics,Zhou2019ODBQC,Qi2019MOSGQQE}, which is closely related to the operator entanglement of the unitary time-evolution operator. One well-studied measure of the operator spreading is the out-of-time ordered correlator (OTOC) \cite{Ponte2015PDEMLQS,Hosur2016Chaos,Garttner2016MOCMQSTQM,Fan2017OTOC,Li2017MOCNMRQS,Keyserlingk2018Operator,Rakovszky2018Diffusive}. We calculate the operator-averaged OTOC for locally-scrambled quantum dynamics using the entanglement feature Hamiltonian. Then from the asymptotic behavior of the OTOC, we determine the butterfly velocity $v_B$, which characterizes the effective Leib-Robinson velocity and the rate of operator entanglement growth\cite{Hartman2013Time,Hosur2016Chaos,Khemani2018Velocity-dependent,Xu2018LQFS,Couch2019The-Speed}. On the other hand, we can also calculate the entanglement velocity $v_E$ from the entanglement dynamics at different volume-law coefficient $s$. By comparing $v_B$ and $v_E$, we check and confirm the previous conjectures about the velocity bounds $v_E\leq (\ln d-|s|)v_B$, as proposed in \refcite{Jonay2018CDOSE,Couch2019The-Speed}. 
		
This paper is organized as follows: in \ref{ansatz}, we outline the \textit{excitation spectrum} of the multi-region entanglement continuum, such as its boundaries and the gap between single-region and multi-region entanglements. This gap $\Delta$ is finite in the thermodynamic limit and contributes to the single-region entanglement dynamics. In this section, we also propose our MPS ansatz for describing the entanglement of generic locally scrambled states as they evolve from product state, through an area law phase eventually thermalizing to a volume law configuration. Despite that the \textit{physical} state's entanglement rapidly grows as it evolves, the corresponding entanglement feature state, which describes the full single and multi-region entanglement structure, is well described by our $D=2$ MPS ansatz. The ansatz enables to calculate exactly many features of this multi-region continuum, including its boundaries and gap, which then illuminates how to interpret our MPS parameters $\alpha,\theta$, whose dynamics are explored in the following section. In \ref{entanglement-dynamics}, we briefly summarize the meaning of locally scrambled quantum dynamics and the EF Hamiltonian ansatz parameters $g,\beta$. To study Haar random and swap circuits using an EF Hamiltonian, we define a continuum limit for these models by analyzing their \textit{entanglement cut} (domain wall) dynamics in the EF Hilbert space. Having defined a continuum limit for these models, we may then focus exclusively on entanglement dynamics generated by the EF Hamiltonian to study locally scrambled circuits. We find that entanglement velocity, and therefore single-region entanglement dynamics, depend on $\Delta$ which is unique to the multi-region continuum, and can in fact dominate the dynamics. In \ref{operator-dynamics}, we derive the butterfly velocity from analyzing the infinite temperature operator averaged OTOC, and show large system size numerics generated by large-$D$ MPS that agree with our result. Lastly, we compare our butterfly velocity with known bounds on entanglement velocity proposed in \refcite{Jonay2018CDOSE,Couch2019The-Speed}, and find that they agree.

\section{Matrix Product State Ansatz for Multi-Region Entanglement}\label{ansatz}

\subsection{Multi-Region Entanglement}

Quantum many-body system can exhibit rich and complex entanglement structures. We consider an isolated many-body system described by a pure state $\ket{\Psi}$, and focus on the bipartite entanglement quantified by the von Neumann or R\'enyi entanglement entropies. In this work, we will restrict to the 2nd R\'enyi entanglement entropy, which will admit the simplest formulation in the entanglement feature approach. Given the many-body state $\ket{\Psi}$, the 2nd R\'enyi entanglement entropy over a region $A$ is defined as
\eq{S_\Psi(A)=-\ln\Tr_A\rho_A^2,}
where $\rho_A=\Tr_{\bar{A}}\ket{\Psi}\bra{\Psi}$ is the reduced density matrix of the subsystem $A$. $S_\Psi(A)$ characterize the amount of entanglement between the entangling region $A$ and its complement $\bar{A}$. Suppose the many-body system is a chain of qudits arranged along a one-dimensional lattice. The region $A$ can be taken to be any subset of the qudits, and does not need to be a continuous segment. However, most current studies of the entanglement dynamics have focused on the growth of entanglement entropy in a single region or the half system. What about the multi-region entanglement entropies? How do they evolve under non-equilibrium quantum dynamics? Do they mutually affect each other during the evolution? These are the problems that we will explore in this work. 

\begin{figure}[htbp]
\begin{center}
\includegraphics[width=0.85\columnwidth]{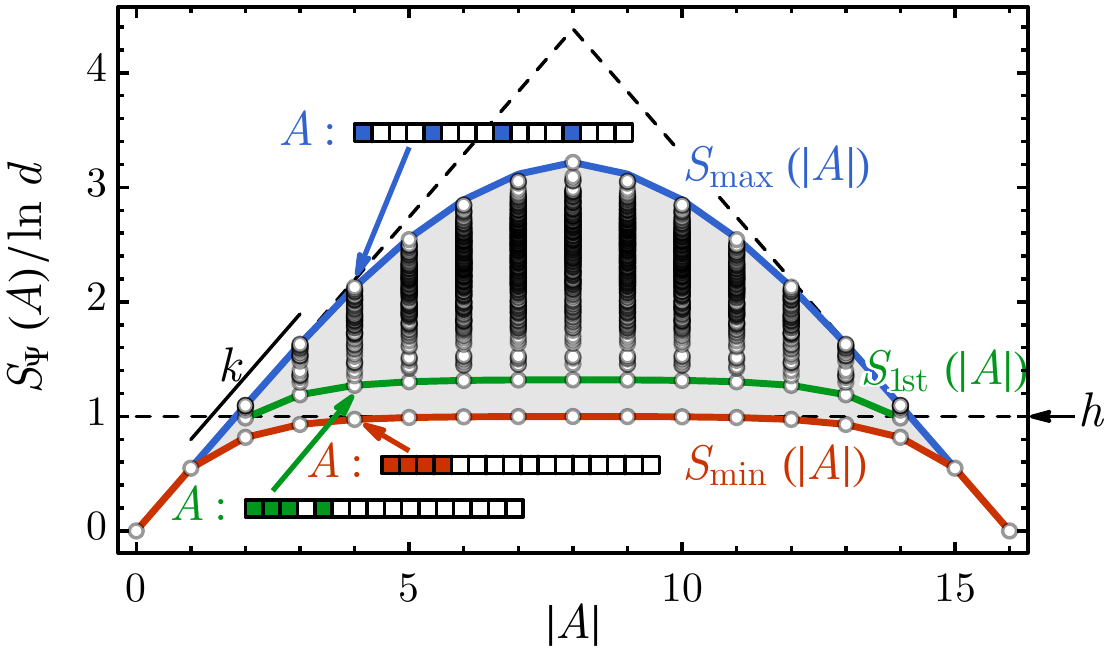}
\caption{Entanglement entropy $S_\Psi(A)$ by the region size $|A|$ for a typical many-body state on a one-dimensional lattice of $N=16$ sites. The $S_\Psi(A)$ data are produced by the $D=2$ MPS model at $(\alpha,\theta)=(\frac{1}{2},\frac{\pi}{4})$ at qudit dimension $d=2$.}
\label{fig: spectrum}
\end{center}
\end{figure}

To study these problems, we first need to organize the multi-region entanglement entropy $S_\Psi(A)$ systematically. If $A$ is a single-region, we can parameterize $A$ by its region size $|A|$ (i.e.~the number of qudits in $A$), because the entanglement entropy $S_\Psi(A)$ will only depend on $|A|$ in the presence of translation symmetry. However for multi-regions, the region size $|A|$ is insufficient to parametrize the region $A$, since $S_\Psi(A)$ will also depend on how $A$ is segmented along the one-dimensional chain. \figref{fig: spectrum} displays the scatter plot of the entanglement entropy $S_\Psi(A)$ with respect to the region size $|A|$ for a typical quantum many-body state. Similar phenomenology derived from a random-matrix framework is discussed in \cite{szyniszewski2019randommatrix}. Data points of $S_\Psi(A)$ distributes in a dome-shaped continuum bounded between the lower edge $S_\text{min}(|A|)$ (in red) and the upper edge $S_\text{max}(|A|)$ (in blue). Given the region size $|A|$, the different values of $S_\Psi(A)$ originates from different segmentations of the entangling region $A$. Because $S_\Psi(A)$ always increases with the number of entanglement cuts in $A$ when $|A|$ is fixed, the lower-bound $S_\text{min}(|A|)$ should be given by the single-region entanglement entropy which has the least number of entanglement cuts, while the upper-bound $S_\text{max}(|A|)$ should correspond to the entangling region $A$ being disjoint sites separated from each other as far as possible which has the most number of entanglement cuts. All the multi-region entanglement entropies are distributed between the curves $S_\text{min}(|A|)$ and $S_\text{max}(|A|)$, forming a dense continuum in the thermodynamic limit $N\to\infty$, which might be dubbed as the \emph{multi-region continuum}.

If we imagine the entanglement entropy $S_\Psi(A)$ as a kind of ``energy'' associated with each region $A$, \figref{fig: spectrum} can be viewed as an ``excitation spectrum'' of entangling regions. It is the whole spectrum that fully characterize the entanglement structure of the underlying quantum many-body state $\ket{\Psi}$. $S_\text{min}(|A|)$ describes how the ``ground state energy'' varies with the region size $|A|$, which is mostly discussed in the literature. But we are also curious about the ``excited states'' in the spectrum. For example, $S_\text{1st}(|A|)$ describes how the ``1st excited state energy'' varies with $|A|$, as the green curve in \figref{fig: spectrum}. The entangling region $A$ that contributes to $S_\text{1st}(|A|)$ always contains two subregions of sizes $(|A|-1)$ and $1$, separated by one site in between, which provides an example of the multi-region entanglement. One may further define the ``excitation gap'' between $S_\text{1st}(|A|)$ and $S_\text{min}(|A|)$ as
\eq{\label{eq: Delta}\Delta(|A|)=S_\text{1st}(|A|)-S_\text{min}(|A|).}
\refcite{Fan2020Self-Organized} found that the \emph{entropy gap} $\Delta(|A|)$ plays an important role in quantifying the error-correcting capacity in sub-thermal volume-law states. This motivates us to further investigate the multi-region entanglement.

However, it is still challenging to organize all the $2^N$ entropies. To meet this challenge, we took the ``entropy-energy correspondence'' farther to define the ``Boltzmann weight'' for each region $A$
\eq{W_\Psi(A)=e^{-S_\Psi(A)},}
which was first introduced as the \emph{entanglement feature} in \refcite{You2018Machine,You2018Entanglement}. In the case of 2nd R\'enyi entropy, the entanglement feature $W_\Psi(A)=\Tr_A\rho_A^2$ is simply the purity of subsystem $A$. If we attempt to arrange the entanglement features as components of a single vector, the number of components ($2^N$) will be the same as that of the state vector of a $N$-spin system. This motivates us to organize the $2^N$ entanglement features into a fictitious many-body state, called the entanglement feature state,\cite{Kuo2019Markovian}
\eq{\label{eq: def W}\ket{W_\Psi}=\sum_{[\sigma]}W_\Psi[\sigma]\ket{[\sigma]},}
where the $2^N$ different entangling regions $A$ are equivalently represented as the $2^N$ Ising configurations $[\sigma]=(\sigma_1,\sigma_2,\cdots,\sigma_N)$ with one-to-one correspondence
\eq{\sigma_i=\left\{\begin{array}{ll}+1\;(\uparrow) & i\notin A,\\
-1\;(\downarrow) & i\in A.\end{array}\right.}
In this way, all regions (no matter single-regions or multi-regions) are treated on equal footing. The entanglement entropy of any region $A$ can be taken back from the entanglement feature state $S_\Psi(A)=-\ln\braket{A}{W_\Psi}$, where $\ket{A}$ is the Ising basis state with down-spins in region $A$ and up-spins in the complement of $A$.

\subsection{Matrix Product State Representation}

At this point, \eqnref{eq: def W} does not seem to really simplify the problem, other than encoding the $2^N$ entanglements into a many-body state. However, an important observation is that the entanglement feature $W_\Psi[\sigma]$ is always positive. According to \refcite{Grover2015Entanglement}, many-body states with positive wavefunctions in a local basis should typically exhibit a constant-law scaling of entanglement entropy, which is only violated in fine-tune cases. Since entanglement feature states $\ket{W_\Psi}$ always have positive components, they should have low entanglement and should admit efficient matrix product state (MPS)\cite{Verstraete2008Matrix} representations. Here we would like clarify that the entanglement feature state $\ket{W_\Psi}$ was originally introduced to describe the entanglement property of the physical quantum state $\ket{\Psi}$, but $\ket{W_\Psi}$ itself as a many-body state also has its own entanglement properties. Given the sign-free nature of $\ket{W_\Psi}$, the entanglement of $\ket{W_\Psi}$ should typically follow a constant-law, regardless of the entanglement properties of corresponding physical state $\ket{\Psi}$, although we have not been able to strictly prove that $\ket{W_\Psi}$ must be constant-law entangled in every cases. Nevertheless, as we will see, even if the underlying physical state $\ket{\Psi}$ is a maximally entangled Page state (i.e.~a random state in the many-body Hilbert space), the corresponding entanglement feature state $\ket{W_\text{Page}}$  still remains constant-law entangled and can be precisely written as a MPS with bond dimension $D=2$. Among all currently known examples, the $\ket{W_\Psi}$ will be most entangled at the entanglement transition\cite{Vasseur2018Entanglement,Skinner2019MPTDE,Li2018QZEMET,Wu2019Entanglement,Li2019METHQC,Szyniszewski2019ETFVWM_PRBVERSION,Choi2019QECEPTRUCWPM,Jian2019MCRQC,Bao2019TPTRUCWM}, where it exhibits a logarithmic-law entanglement. Since it requires fine-tuning to hit the entanglement transition, we argue that for most cases, the entanglement feature state should be MPS-representable.

Therefore, assuming translation symmetry, we propose the following MPS ansatz for the entanglement feature state $\ket{W_\Psi}$
\eq{\label{eq: MPS}W_\Psi[\sigma]\propto\Tr \Big(\prod_i M_\Psi^{\sigma_i}\Big),}
where $M_\Psi^{\sigma_i}$ is a $D\times D$ matrix that depends on $\sigma_i=\pm 1$. We take the periodic boundary condition along the chain, such that the product of matrices can be simply traced over without specifying additional boundary conditions. The right-hand-side of \eqnref{eq: MPS} is not properly normalized yet. The normalization constant should be determined by $\braket{\emptyset}{W_\Psi}=1$, because the entanglement entropy $S_\Psi(\emptyset)=0$ for empty region ($A=\emptyset$) must be zero. Based on the MPS representation, we can apply efficient numerical algorithms, such as the time-evolving block decimation (TEBD), to simulate the entanglement dynamics for large systems. Such numerical approach has been explored in 
\refcite{Fan2020Self-Organized} recently. 

Here we would like to further investigate along the analytic direction. We will construct the minimal MPS model for the entanglement feature state to capture the major features of the multi-region entanglement in \figref{fig: spectrum}. In particular, we will consider the MPS ansatz with bond dimension $D=2$. We could also consider a larger bond dimension for a stronger representation power, but for the purpose of analytical treatment here, we would like to keep the bond dimension as small as possible, such that we can possibly interpret the MPS parameters in the end. The efficacy of the $D=2$ MPS is numerically verified in \appref{app: D=2}, which shows that $D=2$ MPS is already successful in capturing all the multi-region entanglement over the entire thermalization process.

For $D=2$, $M_\Psi^{\sigma}$ will be a $\sigma$-dependent $2\times2$ matrix of the following form
\eq{\label{eq: M}M_{(\alpha,\theta)}^{\sigma}=\cosh \alpha\; I+\sinh\alpha (\sin \theta\; X+\sigma \cos \theta\; Z),}

where $I,X,Z$ denote the identity, Pauli-$x$, and Pauli-$z$ matrices respectively. The ansatz is only controlled by two real parameters $\alpha\geq 0$ and $0\leq \theta\leq\pi/2$. The form in \eqnref{eq: M} can be determined based on the following considerations:
\begin{enumerate}[label=(\roman*)]
\item For pure state $\ket{\Psi}$, the entanglement entropy in region $A$ should be the same as that in the complement region $\bar{A}$, i.e.~$S_\Psi(A)=S_\Psi(\bar{A})$. This implies $W_\Psi[\sigma]=W_\Psi[-\sigma]$ for the entanglement feature, i.e.~the entanglement feature state $\ket{W_\Psi}$ must respect the $\dsZ_2$ symmetry ($\sigma\to-\sigma$). But before imposing the $\dsZ_2$ symmetry on the MPS ansatz, we notice that the matrix $M_{\Psi}^{\sigma}$ carries a gauge freedom, since $\ket{W_\Psi}$ is invariant under the following gauge transformation
\eq{M_{\Psi}^{\sigma}\to G M_{\Psi}^{\sigma}G^{-1},}
induced by any $G\in\mathrm{GL}(2,\dsC)$. Therefore the $\dsZ_2$ symmetry action on $M_{\Psi}^{\sigma}$ will be followed by a corresponding gauge transformation in general\cite{Sanz2009Matrix,Pollmann2010Entanglement,Kull2017Classification}. We can choose the gauge transformation to be $G=X$,\cite{Fan2020Self-Organized} then the $\dsZ_2$ symmetry requires $M_{\Psi}^{\sigma}=XM_{\Psi}^{-\sigma}X$, which can be resolved by $M_{\Psi}^{\sigma}=(c_0 I+c_1 X)+\sigma(\ii c_2 Y+ c_3 Z)$. The coefficients $c_{0,1,2,3}\in\dsR$ should all be real to ensure that the resulting entanglement features are real. 
\item We can always rescale $M_{\Psi}^{\sigma}$ by an overall factor, such as $M_{\Psi}^{\sigma}\to c M_{\Psi}^{\sigma}$. The factors will be absorbed into the normalization constant in \eqnref{eq: MPS}, which can always be fixed by $\braket{\emptyset}{W_\Psi}=1$ in the end. So we are free to set $c_0=1$ by rescaling. 
\item We can use the gauge freedom to eliminate $c_2$ (as a gauge fixing) by performing the gauge transformation of $G=a I+b X$, where $a,b$ can be any solution of $a^2+b^2-2 a b c_3/c_2=0$. 
\item The remaining parameters $c_1$ and $c_3$ can be parametrized by an positive amplitude $c\geq 0$ and an angle $\theta$ following $c_3+\ii c_1= c\,e^{\ii\theta}$, such that $M_{\Psi}^{\sigma}=I+c(\sin\theta X+\sigma \cos\theta Z).$ Using the gauge transformation of $G=Z$ or $G=X$, we can flip the sign of the coefficient in front of $X$ or $Z$ independently. Thus we can make both $\sin \theta$ and $\cos\theta$ positive. So we only need to consider $0\leq \theta\leq \pi/2$. 
\item Now the eigenvalues of $M_{\Psi}^{\sigma}$ are $1\pm c$. To ensure that the entanglement features $W_\Psi[\sigma]$ are positive, we must at least require both eigenvalues to be positive. If, for example, $c>1$ and $\theta=0$, we can show that the single-site entanglement feature will become negative. Thus we should have $0\leq c <1$, so that $c$ can be rewritten as $c=\tanh\alpha$.
\end{enumerate}
Thus we end up with the final form in \eqnref{eq: M} (up to additional rescaling by $\cosh\alpha$).

\subsection{Edges of Multi-Region Continuum}

The MPS ansatz \eqnref{eq: M} provides a minimal model for all the $2^N$ entanglement entropies in terms of two real parameters $(\alpha,\theta)$,
\eq{\label{eq: S}S_{(\alpha,\theta)}[\sigma]=-\ln\Tr\Big(\prod_iM_{(\alpha,\theta)}^{\sigma_i}\Big)+S_0,}
where the background entropy $S_0=\ln(2\cosh \alpha N)$ is attached to ensure the entanglement entropy $S_{(\alpha,\theta)}(\emptyset)=0$ vanishes for empty region (which corresponds to properly normalize the entanglement feature state). \figref{fig: spectrum} is actually generated by \eqnref{eq: S} at $(\alpha,\theta)=(\frac{1}{2},\frac{\pi}{4})$, which demonstrated that the minimal model can describe the multi-region continuum. In particular, we can determine both its lower edge $S_\text{min}(|A|)$ and its upper edge $S_\text{max}(|A|)$, see \appref{app: edge} for derivation. The lower edge is given by the single-region configuration,
\eqs{\label{eq: Smin}S_\text{min}(|A|)&=-\ln\Tr(M^{\downarrow})^{|A|}(M^{\uparrow})^{N-|A|}+S_0\\
&=-\ln\big(\sin^2\theta+\cos^2\theta\tfrac{\cosh \alpha(N-2|A|)}{\cosh \alpha N}\big).}
The upper edge corresponds to the entangling region $A$ of equally spaced single sites. Given the region size $|A|$, the region $A$ will be a disjoint union of $|A|$ sites separated from their neighbors by $N/|A|-1$ sites (see the blue inset of \figref{fig: spectrum}). Ignoring the subtlety of the possible incommensurability between $|A|$ and $N$, the upper edge for $|A|<N/2$ should read
\eqs{\label{eq: Smax}S_\text{max}(|A|)&= -\ln\Tr\big(M^{\downarrow}(M^{\uparrow})^{N/|A|-1}\big)^{|A|}+S_0\\
&=-\ln\frac{\cosh \eta |A|}{\cosh \alpha N},}
where $\eta$ also depends on $|A|$ and is given by $\cosh\eta=\sin^2\theta\cosh\frac{\alpha N}{|A|}+\cos^2\theta\cosh\frac{\alpha(N-2|A|)}{|A|}$.
For $|A|>N/2$, we simply take its $\dsZ_2$ reflection $S_\text{max}(|A|)=S_\text{max}(N-|A|)$.

In the thermodynamic limit $N\to\infty$, \eqnref{eq: Smin} and \eqnref{eq: Smax} reduce to (see \appref{app: edge})
\eqs{\label{eq: S largeN}S_\text{min}(|A|)&=-\ln(\sin^2\theta+\cos^2\theta\;e^{-2\alpha|A|}),\\
S_\text{max}(|A|)&=-\ln(\sin^2\theta+\cos^2\theta\;e^{-2\alpha})|A|.}
Usually when we discuss the area-law v.s.~volume-law entanglement, we are talking about the scaling of the lower edge $S_\text{min}$ with respect to $|A|$. Here we could also talk about the scaling of the upper edge $S_\text{max}$, which, according to \eqnref{eq: S largeN}, is always volume-law (even for the area-law state). Such behavior is rather trivial to understand, as the ``area'' $|\partial A|$ is proportional to the volume $|A|$ in this case. So let us focus on the lower edge, which generally exhibits an area-law scaling (unless $\theta=0$), as $S_\text{min}$ gradually saturates to $-2\ln\sin\theta$ for large $|A|$. Only when $\theta=0$, the lower edge also exhibits a volume-law scaling and coincides with the upper edge, i.e.~$S_\text{min}(|A|)=S_\text{max}(|A|)=2\alpha |A|$, such that the multi-region continuum is squeezed to vanish in this case. However, we should mention that the vanishing multi-region continuum for all volume-law states is an artifact of the $D=2$ MPS model. When a non-vanishing multi-region continuum appears on top of the volume-law lower edge $S_\text{min}$, it implies that the volume-law state has a non-trivial quantum error correcting capacity, which has been discussed in \refcite{Choi2019QECEPTRUCWPM,Fan2020Self-Organized}. As a minimal model, the $D=2$ MPS ansatz has too few parameters to capture the error-correcting properties in the volume-law state. If we extend the MPS ansatz to higher bond dimension $D>2$, the error-correcting volume-law state can be described as well (as has been shown in \refcite{Fan2020Self-Organized}).

\subsection{Physical Meaning of MPS Parameters}

\begin{figure}[htbp]
\begin{center}
\includegraphics[width=0.82\columnwidth]{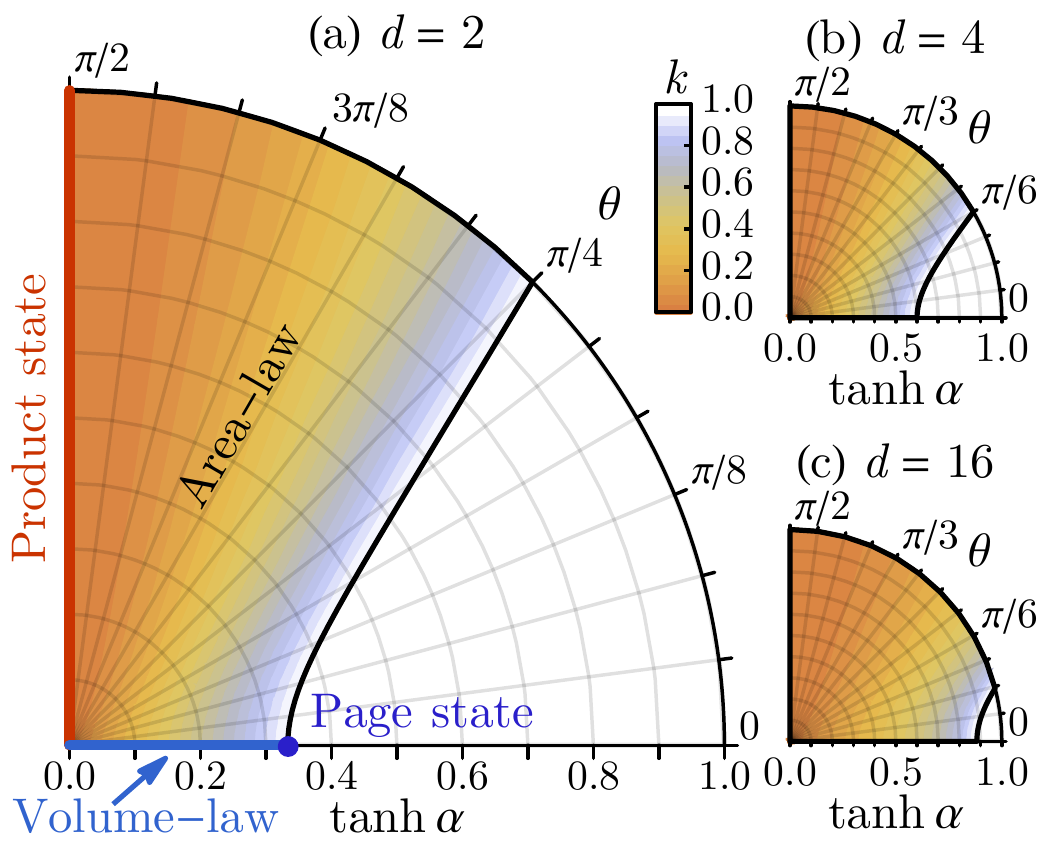}
\caption{Feasible domain of MPS parameters $(\alpha,\theta)$ at different qudit dimensions: (a) $d=2$, (b) $d=4$, (c) $d=16$. The background color indicates the upper edge volume-law coefficient $k$.}
\label{fig: domain}
\end{center}
\end{figure}

To describe the overall shape of the multi-region continuum, we define the following two characteristic quantities (as illustrated in \figref{fig: spectrum}): the area-law plateau height of the lower edge $S_\text{min}(|A|)$
\eq{h\equiv \frac{S_\text{min}(N/2)}{\ln d}=-\log_d \sin^2\theta,}
and the volume-law coefficient of the upper edge $S_\text{max}(|A|)$
\eqs{k&\equiv \lim_{|A|\to0}\frac{\partial_{|A|}S_\text{max}(|A|)}{\ln d}\\
&=-\log_d(\sin^2\theta+\cos^2\theta\;e^{-2\alpha}),}
where $\log_d$ denotes the logarithm of base $d$ (and $d$ is the qudit dimension). The other way round, the MPS parameters $(\alpha,\theta)$ can also be expressed in terms of the height $h$ and the slope $k$ as
\eqs{\label{eq: para}\alpha&=\frac{1}{2}\ln\frac{d^h-1}{d^{h-k}-1},\\
\theta&=\arcsin d^{-h/2}.}
Now the physical meaning of the MPS parameters becomes clear. The $\theta$ parameter is directly determined by the area-law plateau height $h$. For product states, $h=0$, hence $\theta=\pi/2$. For generic area-law states, $h\geq 0$ is positive, hence $0<\theta\leq\pi/2$. For volume-law state, $h\to\infty$ (the area-law plateau never appears), hence $\theta=0$. We summarize these cases in \figref{fig: domain}(a). Once $\theta$ is fixed, the remaining parameter $\alpha$ will be set by the slope $k$ of the upper edge. In particular, for the volume-law state ($\theta=0$), $\alpha$ is directly related to the volume-law coefficient $k$ by $\alpha=\frac{1}{2}k\ln d$, which is half of the entanglement entropy of a single qudit. 

As the volume-law coefficient approaches unity ($k=1$), the system reaches the maximally entangled Page state $(\alpha,\theta)=(\frac{1}{2}\ln d,0)$, where the MPS ansatz in \eqnref{eq: M} reduces to $M_\text{Page}^\sigma=I+\frac{d-1}{d+1}\sigma Z$, which gives the exact description of the Page state entanglement features $\ket{W_\text{Page}}$. Although the underlying physical state $\ket{\Psi}$ is highly entangled, its entanglement features can still be captured by a low-entanglement MPS efficiently, because it generally takes much less entanglement resources to describe the entanglement property of a many-body state $\ket{\Psi}$ than the state itself. Therefore using the MPS approach, we can achieve a huge compression of the many-body entanglement features. Just starting from two characteristic quantities $h$ and $k$, the MPS ansatz \eqnref{eq: S} can provide a comprehensive modeling of entanglement entropies in all the $2^N$ possible regions (see \figref{fig: spectrum}), demonstrating the prediction power of the MPS model. So, within the MPS model, the question of how entanglement entropies evolve in different regions boils down to how MPS parameters (such as $\alpha,\theta$) evolve, which will be investigated in more details soon. 

Before discussing the entanglement dynamics, let us first determine the feasible domain of the MPS parameters. A non-trivial constraint comes from the requirement that the volume-law coefficient $k\leq 1$ must not exceed one, because the entanglement entropy of a single qudit can not be greater than $\ln d$. According to \eqnref{eq: para}, $k\leq 1$ implies
\eq{(d \cos 2\theta+1)\tanh\alpha\leq d-1.}
This inequality further restricts the primitive domain of $\alpha\geq 0$ and $0\leq \theta\leq \pi/2$, leading to the feasible domain shown in \figref{fig: domain}. The shape of the feasible domain varies with the qudit dimension $d$, as illustrated in \figref{fig: domain}(b,c). In the following section, we will study the evolution of the entanglement feature state under locally scrambled quantum dynamics. Using the MPS ansatz developed in this section, we will be able to represent the evolution in the MPS parameter space, which will provide an intuitive picture of how the quantum system thermalizes with time.

\section{Entanglement Dynamics in Locally Scrambled Quantum Systems}\label{entanglement-dynamics}

\subsection{Entanglement Feature Formalism}

Let us put aside the MPS model shortly and discuss the dynamics of generic entanglement feature state $\ket{W_\Psi}$. In the entanglement feature formalism, $\ket{W_\Psi}$ captures the entanglement entropies over all regions for a given quantum many-body state $\ket{\Psi}$. As the state $\ket{\Psi}$ evolves in time, so does its corresponding entanglement feature state $\ket{W_\Psi}$. \refcite{Kuo2019Markovian} has proven that the entanglement dynamics will be governed by an imaginary-time Schr\"odinger equation
\eq{\label{eq: Schrodinger}-\partial_t\ket{W_\Psi}=H_\text{EF}\ket{W_\Psi},}
if the underlying quantum dynamics is \emph{locally scrambled}. A quantum dynamics $\ket{\Psi}\to U\ket{\Psi}$ is said to be locally scrambled, if each step of the unitary evolution $U$ is drawn from a ensemble whose probability measure $P(U)$ is invariant under local basis transformations, i.e.~$P(U)=P(VUV^\dagger)$ for all $V=\prod_i V_{i}$ with arbitrary $V_{i}\in\U(d)$ on site-$i$. Examples of locally scrambled quantum system include random unitary circuits\cite{Nahum2017Quantum,Zhou2019Emergent,Keyserlingk2018Operator,Nahum2018Dynamics}, random Hamiltonian dynamics\cite{Vijay2018Finite-Temperature,You2018Entanglement}, and quantum Brownian dynamics\cite{Lashkari2013TFSC,Xu2018LQFS,Gharibyan2018ORMBSS,Zhou2019ODBQC}. Because local basis information is fully scrambled at each step of the time-evolution, the entanglement dynamics will be Markovian, which can be described by the entanglement feature Hamiltonian $H_\text{EF}$ following \eqnref{eq: Schrodinger}.

As derived in \refcite{Kuo2019Markovian}, up to the nearest neighbor coupling on a one-dimensional lattice, $H_\text{EF}$ should take the following general form (as a many-body spin model)
\eq{\label{eq: HEF}H_\text{EF}=g\sum_{\langle ij\rangle}\frac{1-Z_iZ_j}{2}e^{-\delta(X_i+X_j)-\beta X_i X_j},}
so as to preserve the $\dsZ_2$ symmetry and the normalization of $\ket{W_\Psi}$ under the entanglement dynamics and to respect the time-reversal symmetry. Here $X_i$ and $Z_i$ are Pauli operators acting on site-$i$. The parameter $\delta$ is fixed by the qudit dimension $d$ via $\tanh \delta=1/d$, and $H_\text{EF}$ is only controlled by two free parameters $g$ and $\beta$. The parameter $g$ sets the time-scale and determines how fast the dynamics will happen on the overall scale. The parameter $\beta$ is tied to the type of the quantum dynamics. A few examples are listed in \tabref{tab: beta}. Unlike conventional spin models, $H_\text{EF}$ contains a projection operator $\frac{1-Z_iZ_j}{2}$ on each bond, which imposes dynamic constraints on the motion of domain walls\cite{De-Tomasi2019Dynamics,Yang2019Hilbert-Space}. We will discuss its effect on the entanglement dynamics later.

\begin{table}[htp]
\caption{Examples of locally scrambled quantum dynamics and their corresponding $\beta$ value (in terms of $\tanh\beta$).}
\begin{center}
\begin{tabular}{cc}
Type of quantum dynamics & $\tanh\beta$\\
\hline
Quantum Brownian dynamics & $0$\\
Random unitary circuit (continuum limit) & $1/d^2$\\
Fractional swap circuit (continuum limit, $x\to1$) & $1$
\end{tabular}
\end{center}
\label{tab: beta}
\end{table}

\subsection{Locally Scrambled Quantum Dynamics}

Let us briefly review some important examples of locally scrambled quantum dynamics, in order to gain some intuition about the parameter $\beta$. The first example is the quantum Brownian dynamics (also known as the Brownian random circuit)\cite{Lashkari2013TFSC}, where the quantum many-body state $\ket{\Psi}$ evolves under a time-dependent random Hamiltonian following $\ket{\Psi}\to e^{-\ii H_t \dd t}\ket{\Psi}$, with
\eq{H_t=\sum_{\langle ij\rangle}J_{t,ij}^{ab}T_i^a T_j^b,}
where $T_i^a$ (for $a=1,2,\cdots,d^2$) are $\U(d)$ generators on site $i$ with the normalization $\Tr T_i^{a\dagger}T_i^b=\delta^{ab}$, and the couplings $J_{t,ij}^{ab}$ are independently drawn for each time $t$ and indices $i,j,a,b$ from a Gaussian distribution with zero mean and $d^{-2}$ variance. According to \refcite{Kuo2019Markovian}, the corresponding entanglement dynamics (the evolution of $\ket{W_\Psi}$) is precisely described by \eqnref{eq: HEF}, with $g=2(1-d^{-2})$ and $\beta=0$.

\begin{figure}[htbp]
\begin{center}
\includegraphics[width=0.78\columnwidth]{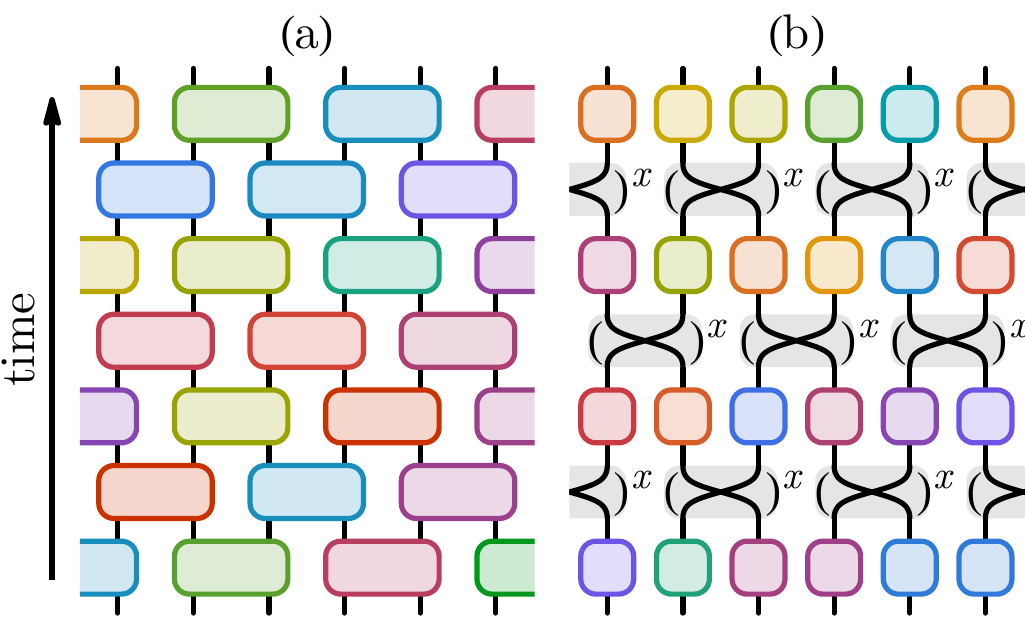}
\caption{(a) Random unitary circuit. (b) Fractional swap circuit. Each color block represents an independent Haar random unitary gate. The fractional swap gate is depicted as swap gate to the fractional power $x$.}
\label{fig: circuit}
\end{center}
\end{figure}

The next example is the random unitary circuit\cite{Nahum2017Quantum}, depicted in \figref{fig: circuit}(a). The physical quantum state $\ket{\Psi}\to U_t\ket{\Psi}$ evolves by the application of random unitary gates layer-by-layer ($U_t$ denotes the whole layer of gates on an equal-time slice at time $t$), where each gate is independently drawn from two-qudit Haar random unitary ensemble. As a quantum circuit model, the time is discrete (each layer is a step in time) and the corresponding entanglement dynamics is described by the following transfer matrix\cite{Kuo2019Markovian}
\eqs{\label{eq: T Haar}\ket{W_\Psi}&\to\prod_{\langle ij\rangle\in\Lambda_{\pm}}T_{ij}^\text{Haar}\ket{W_\Psi},\\
T_{ij}^\text{Haar}&=1-\tfrac{1-Z_iZ_j}{2}\big(1-\tfrac{d}{d^2+1}(X_i+X_j)\big),}
where $\Lambda_{+}$ ($\Lambda_{-}$) denotes the collection of even (odd) bonds which are chosen alternately following the brick-wall pattern of the circuit. Another circuit model is the fractional swap circuit, as illustrated in \figref{fig: circuit}(b), which was first introduced in \refcite{Kuo2019Markovian}. It is constructed by a layer of fractional swap gates (i.e.~swap gates to the fractional power $x\in[0,1]$: $\mathsf{SWAP}^x=\frac{1+e^{\ii x\pi}}{2}+\frac{1-e^{\ii x\pi}}{2}\mathsf{SWAP}$) followed by a layer of on-site Haar random unitary gates, and so on. The corresponding entanglement dynamics is described by the following transfer matrix\cite{Kuo2019Markovian}
\eqs{\label{eq: T SWAPx}\ket{W_\Psi}&\to\prod_{\langle ij\rangle\in\Lambda_{\pm}}T_{ij}^{\mathsf{SWAP}^x}\ket{W_\Psi},\\
T_{ij}^{\mathsf{SWAP}^x}&=1-\tfrac{1-Z_iZ_j}{2}(u-v(X_i+X_j)+w X_iX_j),}
where $(u,v,w)=(d^2 a-b,d a-d b,a-d^2 b)/(d^2-1)$ with $a=(2-\sin^2\frac{x\pi}{2})\sin^2\frac{x\pi}{2}$ and $b=\sin^4\frac{x\pi}{2}$. In the limit of $x\to1$, the fractional swap gate reduces to the swap gate and $(u,v,w)\to(1,0,-1)$, such that the transfer matrix takes a simpler form
\eq{\label{eq: T SWAP}T_{ij}^{\mathsf{SWAP}}=1-\tfrac{1-Z_iZ_j}{2}(1-X_iX_j).}
This limit may be called the swap circuit model, which was proposed\cite{Kuo2019Markovian} to mimic the entanglement dynamics in integrable conformal field theories (CFT) where entanglement spreads with the ballistic propagation of quasi-particles\cite{Calabrese2005EEEOS,Nie2018Signature,Kudler-Flam2019Quantum}.

\subsection{Causal Structure and Continuum Limit}\label{sec: causal}

To make connection to the continuous-time entanglement dynamics described by $H_\text{EF}$, we would like to take the continuum limit of the transfer matrixes $T^\text{Haar}$ and $T^{\mathsf{SWAP}}$. In general, it is unclear how to define the continuum limit for discrete circuit models, but for the random unitary circuit and the swap circuit dynamics, we can use their definitive causal structure to pin down the corresponding parameter $\beta$ in $H_\text{EF}$, hence obtaining the continuum version of their entanglement dynamics.

\begin{figure}[htbp]
\begin{center}
\includegraphics[width=0.9\columnwidth]{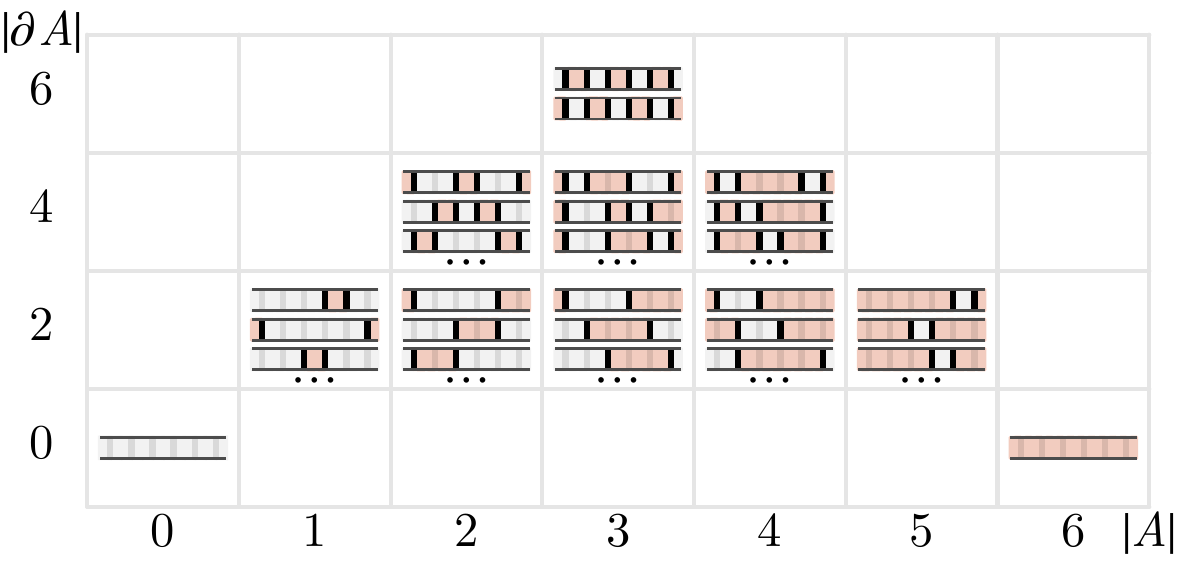}
\caption{Classification of entanglement regions $A$ by their volumes $|A|$ and areas $|\partial A|$ on a $N=6$ lattice.}
\label{fig: regions}
\end{center}
\end{figure}

To reveal the structure of the entanglement feature Hilbert space spanned by the entanglement region basis states $\ket{A}$, we classify the entanglement region $A$ by its volume $|A|$ (the number of sites in $A$) and its area $|\partial A|$ (the number of entanglement cuts, also twice of the number of entanglement regions). In this way, the Hilbert space is partitioned into sectors labeled by $(|A|,|\partial A|)$, as tabulated in \figref{fig: regions}. We would like to understand how the entanglement feature transfer matrix connects different sectors of the Hilbert space. Since the entanglement dynamics is $\dsZ_2$ symmetric, we only need to keep track of the $\dsZ_2$ invariant objects, which are the entanglement cuts (domain walls of Ising spins). Therefore the multi-region entanglement dynamics is essentially a many-body dynamics of entanglement cuts.

\begin{figure}[htbp]
\begin{center}
\includegraphics[width=0.76\columnwidth]{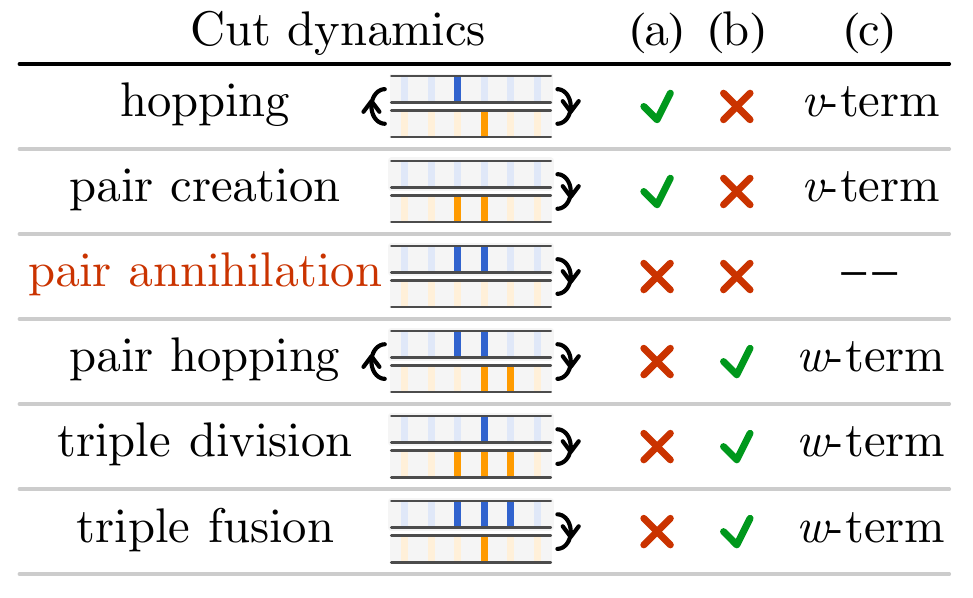}
\caption{Dynamical constraints on the entanglement cut motion in (a) random unitary circuit $T^\text{Haar}$, (b) swap circuit $T^{\mathsf{SWAP}}$. (c) Corresponding terms in $H_\text{EF}$ that leads to the cut dynamics.}
\label{fig: dynamics}
\end{center}
\end{figure}

\begin{figure}[htbp]
\begin{center}
\includegraphics[width=0.78\columnwidth]{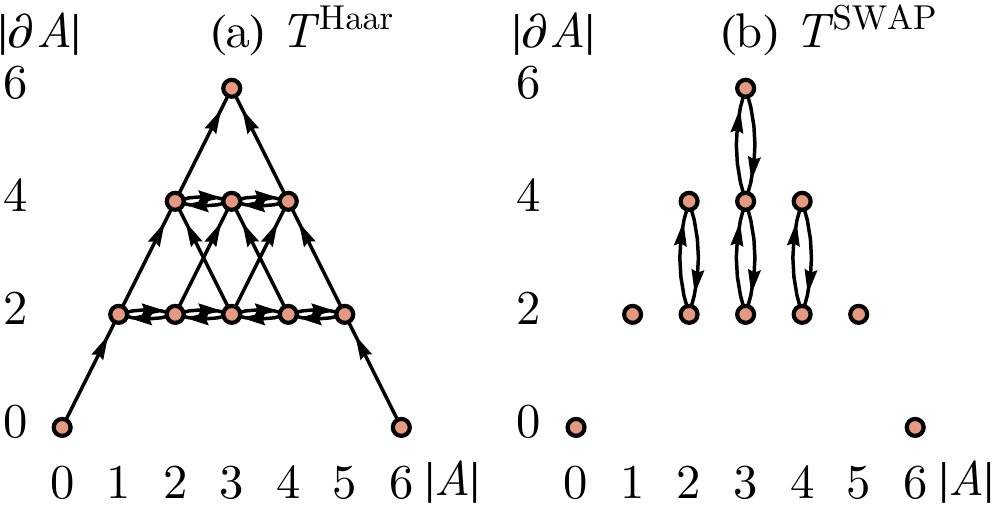}
\caption{Causal structure in the entanglement feature Hilbert space for (a) random unitary circuit, (b) swap circuit. Each note represents a $(|A|,|\partial A|)$-sector. The entanglement features in one sector will only affect those in the other sector along the causal flow (indicated by arrows).}
\label{fig: graph}
\end{center}
\end{figure}

Given the transfer matrix in \eqnref{eq: T Haar} or \eqnref{eq: T SWAP}, the basic moves of  entanglement cuts can be read out and summarized in \figref{fig: dynamics}. The column (a) in \figref{fig: dynamics} indicates whether a process is allowed (checked) or forbidden (crossed) by the transfer matrix $T^\text{Haar}$ for the random unitary circuit defined in \eqnref{eq: T Haar}. The entanglement cuts are free to move  along the lattice and can be created in pairs. But once created, they can not be annihilated, as the pair annihilation process is forbidden. As a result of this dynamical constraint, the entanglement features in the lower-$|\partial A|$ sectors can only affect those in the higher-$|\partial A|$ sectors (via pair creation), but not the other way round, as shown in \figref{fig: graph}(a). In particular, the single-region entanglement is not affected by the multi-region entanglement under the entanglement dynamics, which is a unique property of the random unitary circuits.

Similar analysis can be done for the swap circuit, described by the transfer matrix $T^\mathsf{SWAP}$ in \eqnref{eq: T SWAP}. As shown in \figref{fig: dynamics}(b), a single entanglement cut can not move but a pair of them can hop together; moreover, pairs of entanglement cuts can only be created or annihilated in the presence of the third cut via the triple division or fusion process. One can see that these moves conserve the region volume $|A|$, so the entanglement dynamics only happens within each $|A|$ sector separately, as illustrated in \figref{fig: graph}(b).

The causal structures in \figref{fig: graph} can be formulated as the following algebraic conditions,
\eqs{P_{|\partial A|\leq2k}T^\text{Haar}P_{|\partial A|>2k}=0&\quad(k=0,1,\cdots,N/2),\\
[P_{|A|=n}, T^\mathsf{SWAP}]=0&\quad (n=0,1,\cdots,N),}
where $P_{\cdots}$ denotes the projection operator that projects to the subspace specified by its subscript. We expect their corresponding continuum limit to respect the same causal structure. We assume that the continuous-time entanglement dynamics will be described by the entanglement feature Hamiltonian in \eqnref{eq: HEF}, which can be equivalently written as the following expanded form
\eq{\label{eq: HEF uvw}
H_\text{EF}=\sum_{\langle ij\rangle}\frac{1-Z_iZ_j}{2}(u-v(X_i+X_j)+w X_i X_j),}
with parameters $u,v,w$ related to $g,\beta$ as
\eq{\label{eq: uvw}\left(\begin{array}{c}u\\v\\w\end{array}\right)=\frac{g\cosh\beta}{d^2-1}\left(\begin{array}{c}d^2-\tanh\beta\\d-d\tanh\beta\\1-d^2\tanh\beta\end{array}\right).}
The dynamic process associated to $v$ and $w$ terms are listed in \figref{fig: dynamics}(c). For random unitary circuit, the causal structure requires
\eq{P_{|\partial A|\leq2k}H_\text{EF}P_{|\partial A|>2k}=0\quad(k=0,1,\cdots,N/2),}
which implies $w=0$ and hence $\tanh\beta=1/d^2$. For swap circuit (as the $x\to1$ limit of the fractional swap circuit), the causal structure requires
\eq{[P_{|A|=n}, H_\text{EF}]=0\quad (n=0,1,\cdots,N),}
which implies $v=0$ and hence $\tanh\beta=1$. These parameter correspondences are also obvious by comparing the dynamical constraints between \figref{fig: dynamics}(a,b) and \figref{fig: dynamics}(c). By comparing the $u,v,w$ parameters in \eqnref{eq: T SWAPx} and those in \eqnref{eq: uvw}, we conjecture that $\tanh\beta=\sin^2\frac{x\pi}{2}/(2-\sin^2 \frac{x\pi}{2})$ for the fractional swap circuit at fraction $x$. 

In conclusion, we define the continuum limit of random circuit models according to their definitive causal structures. The results are summarized in \tabref{tab: beta}, which provides us some intuitions about the meaning of $\beta$. However, as a free parameter in $H_\text{EF}$, $\beta$ can take any real positive value in general. We are not yet clear what will be (or how to define) the corresponding microscopic circuit models beyond the examples listed in \tabref{tab: beta}, but this might also be an advantage of the entanglement feature formalism, which allows us to investigate the universal behavior of entanglement dynamics at a higher level without much knowledge about underlying model details.

\subsection{Flow of MPS Parameters}

Having clarified the entanglement feature formalism $-\partial_t \ket{W_\Psi}=H_\text{EF}\ket{W_\Psi}$ and specified the entanglement feature Hamiltonian $H_\text{EF}$ in \eqnref{eq: HEF}, we set out to solve the Schr\"odinger equation to study the entanglement dynamics. However, as a many-body problem, the Schr\"odinger equation is hard to  solve analytically. \refcite{Kuo2019Markovian} has analyzed the long-time asymptotic behavior of the entanglement dynamics (correspond to the ``low-energy'' physics of $H_\text{EF}$), where it was found that all systems will thermalize to the Page state under locally scrambled quantum dynamics with the thermalization (relaxation)  time $\tau$ given by $\tau^{-1}=2g\cosh\beta(1-1/d)^2$. Here we would like to pursuit a different direction by representing the entanglement feature state $\ket{W_\Psi}$ as an MPS and exploring the flow of MPS parameters induced by the entanglement dynamics. This is in line with the MPS-based time dependent variational principle (TDVP)\cite{Haegeman2011Time-Dependent,Haegeman2016Unifying,Leviatan2017Quantum,Kloss2018Time-dependent,Goto2019Performance} developed to simulate quantum many-body dynamics.

\begin{figure}[htbp]
\begin{center}
\includegraphics[width=0.98\columnwidth]{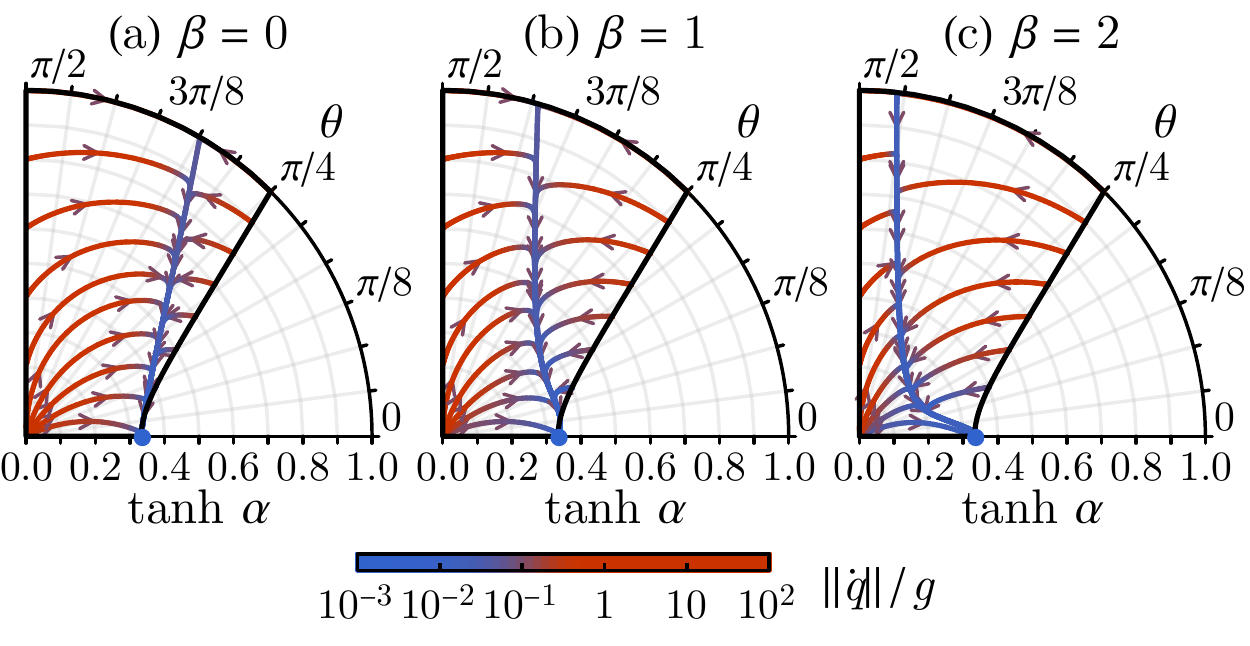}
\caption{Flow of MPS parameters under the entanglement dynamics at (a) $\beta=0$, (b) $\beta=1$, (c) $\beta=2$, in the thermodynamic limit $N\to\infty$ with $d=2$ qudit dimension. Curve color indicates the flow speed.}
\label{fig: flow}
\end{center}
\end{figure}

The classical dynamics of the MPS parameter $q\equiv(\alpha,\theta)$ (unified as $q$) follows from (see \appref{app: MPS} for derivation)
\eq{\label{eq: MPS dyn}-\bra{\partial_{q_i}W_\Psi}W_\id^{-1}\ket{\partial_{q_j}W_\Psi}\dot{q}_j=\bra{\partial_{q_i}W_\Psi}W_\id^{-1}H_\text{EF}\ket{W_\Psi},}
where $\ket{W_\Psi}$ is given by the MPS ansatz in \eqnref{eq: MPS} and \eqnref{eq: M}, $H_\text{EF}$ is given by \eqnref{eq: HEF}. The operator $W_\id^{-1}=(\tanh\delta\sinh\delta)^N e^{-\delta\sum_i X_i}$ is a non-trivial metric that maps a state $\ket{W_\Psi}$ to its dual state $\bra{W_\Psi}W_\id^{-1}$ in the dual Hilbert space (since the entanglement feature Hamiltonian $H_\text{EF}$ is not Hermitian, the state vector and its dual vector do not coincide)\footnote{See \refcite{Kuo2019Markovian} for the meaning of the notation $W_\id^{-1}$ and why it should correspond to the metric in the entanglement feature Hilbert space.}. By solving \eqnref{eq: MPS dyn} numerically (see \appref{app: MPS} for details), we can map out the flow of MPS parameters in the parameter space as shown in \figref{fig: flow} for different types of dynamics specified by the $\beta$ parameter in $H_\text{EF}$.

\begin{figure}[htbp]
\begin{center}
\includegraphics[width=0.95\columnwidth]{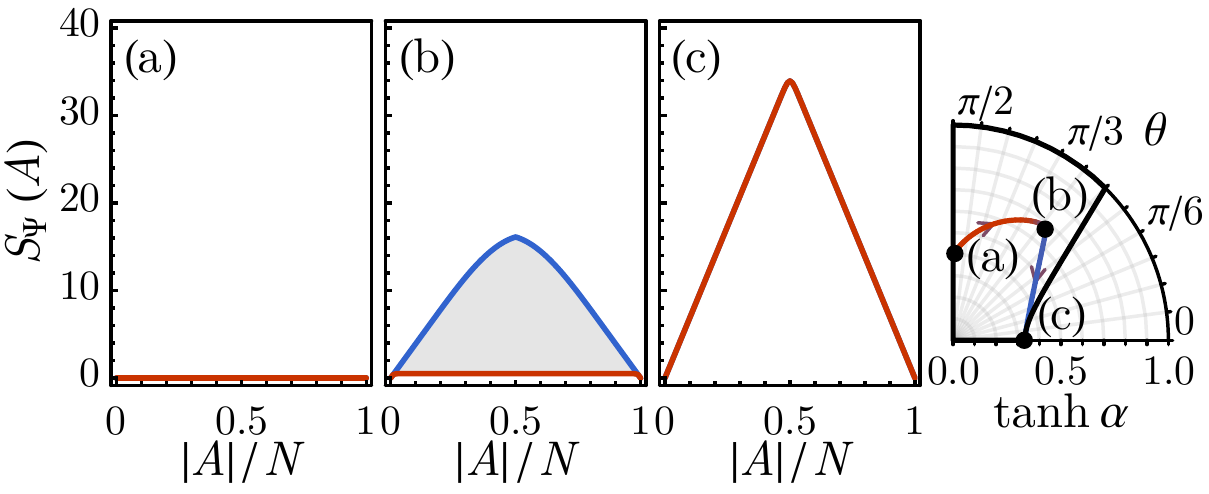}
\caption{Evolution of the entanglement entropy continuum along a typical trajectory in the MPS parameter space through (a) the initial product state, (b) the intermediate area-law state, and (c) the final Page state. The red (blue) curves outlines the bottom $S_\text{min}$ and top $S_\text{max}$ edges of the continuum, which coincide in (a,c). (a-b) the early stage and (b-c) the late stage evolutions are distinct.}
\label{fig: samples}
\end{center}
\end{figure}

As one can see in \figref{fig: flow}, for various choice of the $\beta$ parameter, the MPS parameters all flow to the universal fixed point $(\alpha,\theta)=(\frac{1}{2}\ln d,0)$ which corresponds to the Page state. But the evolution typically divides into two distinct stages. In the early stage (in red), the parameters quickly converge to a mainstream curve. The process is characterized by a large flow speed $\|\dot{q}\|$. Then in the late stage (in blue), the parameters slowly flow along the mainstream towards the Page state fixed point, with a flow speed $\|\dot{q}\|$ orders of magnitude smaller than that of the early stage. The mainstream curve (the central blue curve in \figref{fig: flow}) exhibits a systematic dependence on $\beta$. We recover the multi-region continuum of entanglement entropies from the MPS parameters $(\alpha,\theta)$ along the evolution trajectory, as illustrated in \figref{fig: samples}. Starting from a product state \figref{fig: samples}(a), where $S_\Psi(A)=0$ in all regions (such that $S_\text{max}=S_\text{min}=0$). Under the early stage evolution, the upper edge $S_\text{max}$ quickly evolves to a volume-law curve, while the lower edge $S_\text{min}$ remains almost the same near zero, which establishes the multi-region continuum between $S_\text{min}$ and $S_\text{max}$ as in \figref{fig: samples}(b). Then the evolution enters the late stage, where the area-law lower edge $S_\text{min}$ gradually catches up and finally approaches the (volume-law) Page curve together with $S_\text{max}$ as in \figref{fig: samples}(c). We can thus identify the early (late) stage evolution with the local (global) thermalization.

\begin{figure}[htbp]
\begin{center}
\includegraphics[width=0.88\columnwidth]{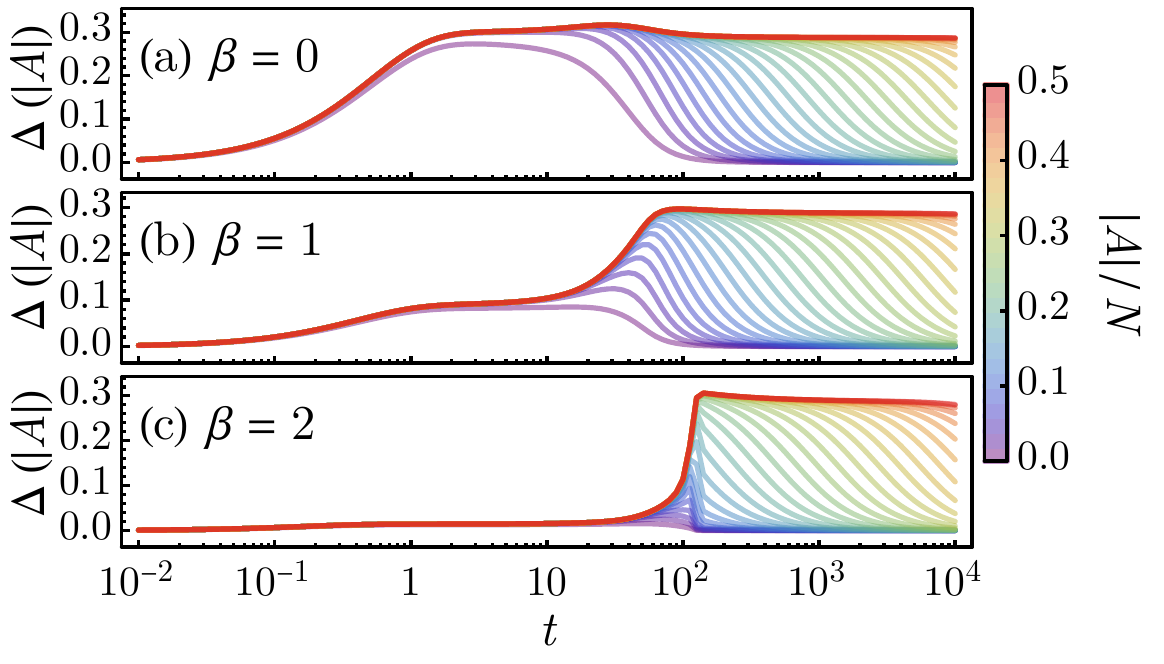}
\caption{Time-evolution of the entropy gap $\Delta(|A|)$ for different dynamics at (a) $\beta=0$, (b) $\beta=1$, (c) $\beta=2$ (assuming $g=1$). Curves of different colors correspond to different $|A|/N$ ratio.}
\label{fig: gap}
\end{center}
\end{figure}

To characterize the two evolution stages more quantitatively, we investigate the behavior of the entropy gap $\Delta(|A|)=S_\text{1st}(|A|)-S_\text{min}(|A|)$, which was introduced in \eqnref{eq: Delta}. Using the MPS representation of the entanglement feature state, the ``1st excited'' entanglement entropy $S_\text{1st}(|A|)$ is given by
\eq{S_\text{1st}(|A|)=-\ln\Tr(M^\downarrow)^{|A|-1}M^\uparrow M^\downarrow (M^\uparrow)^{N-|A|-1}+S_0,}
while $S_\text{min}(|A|)$ was given in \eqnref{eq: Smin}. By definition, the entropy gap is positive, i.e.~$\Delta(|A|)\geq 0$. It characterizes the minimal deviation of  the multi-region entanglement from the single-region entanglement, which reflects the prominence of the multi-region entanglement. In both the short-time $t\to 0$ and long-time $t\to \infty$ limits, the entropy gap vanishes $\Delta(|A|)=0$ as the whole multi-region continuum collapses. During the time-evolution, as shown in \figref{fig: gap}, the entropy gap $\Delta(|A|)$ first increases in the early stage as the multi-region continuum gets established, and then decreases in the late stage as $S_\text{min}$ catches up with $S_\text{max}$ and wipes out the continuum. The decrease of $\Delta(|A|)$ happens gradually from small region to large region as thermalization progresses in the system. 

In conclusion, by studying the evolution of MPS parameters, we effectively parametrize the evolution of all the multi-region entanglement entropies approximately. We find that the two-stage evolution is universal for all types of locally scrambled quantum dynamics, however dynamic details are different for different types of dynamics as specified by different $\beta$ values.

\subsection{Effect of Multi-Region Entanglement on Entanglement Velocity}

Despite of the concise and effective MPS description of the multi-region entanglement on the overall level, it is still not transparent how the multi-region entanglement affects the entanglement dynamics and in what circumstance is it important to consider the multi-region entanglement. We will analyze these problems from the perspective of single-region entanglement dynamics. 

As discussed in \secref{sec: causal}, entanglement entropies from different $|\partial A|$-sectors are causally related under the entanglement dynamics, so it is generally not possible to write down a close-form equation for the evolution of the single-region entanglement entropy. Only in the special case of $\tanh\beta=1/d^2$ (corresponding to the random unitary circuit), the single-region entanglement is not affected by the multi-region entanglement, such that a close-form equation within the single-region sector ($|\partial A|=2$) becomes possible. Earlier  works\cite{Jonay2018CDOSE,Zhou2019Emergent,Zhou2019Emergent} that formulated the single-region entanglement dynamics are indeed based on the random unitary circuit model (either explicitly or implicitly). Once we deviate from this special point, the multi-region entanglement will enter the dynamic equation for the single-region entanglement. We will explore its effect in the following.

Our starting point is the entanglement feature formalism, where the entanglement entropy is given by $S_\Psi(A)=-\ln\braket{A}{W_\Psi}$ with $\ket{A}$ being the basis state labeled by the entanglement region $A$. From $-\partial_t\ket{W_\Psi}=H_\text{EF}\ket{W_\Psi}$, we have
\eq{\label{eq: dS}\partial_t S_\Psi(A)=\frac{\bra{A}H_\text{EF}\ket{W_\Psi}}{\braket{A}{W_\Psi}}.}
Due to the dynamic constrained imposed by the projection operator $\frac{1-Z_iZ_j}{2}$ in the entanglement feature Hamiltonian $H_\text{EF}$, when $H_\text{EF}$ acting on $\bra{A}$ from the right, only those terms across the entanglement cuts of the region $A$ are active. So if the number of entanglement cuts $|\partial A|$ is small, the corresponding dynamics will be simple. The simplest non-trivial case is the dynamics of the single-region entanglement entropy $S_\text{min}(|A|)$. Applying the general formula \eqnref{eq: dS} to $S_\text{min}(|A|)$, the single-region entanglement growth follows
\eqs{\label{eq: dSmin}\partial_t S_\text{min}(|A|)&=2\Gamma\big(\partial_{|A|}S_\text{min}(|A|)\big)+2\Omega\big(\Delta(|A|)\big),\\
\Gamma(s)&=ge^{-\beta}\frac{d^2+1}{d^2-1}\Big(1-\frac{2d}{d^2+1}\cosh s\Big),\\
\Omega(\Delta)&=\frac{g\cosh\beta}{d^2-1}(d^2 \tanh\beta-1)(1-e^{-\Delta}).}
The first term $\Gamma(s)$ describes the dynamics within the single-region sector, which depends on the single-region entanglement entropy gradient $s(|A|)\equiv\partial_{|A|}S_\text{min}(|A|)$. The second term $\Omega(\Delta)$ describes the contribution of multi-region entanglement, which depends on the entropy gap $\Delta(|A|)$ introduced previously in \eqnref{eq: Delta}. The factor 2 in front of both terms comes from the two entanglement cuts associated with a single-region $A$. The effect of the multi-region entanglement on the single-region entanglement dynamics enters explicitly from the $\Omega(\Delta)$ term, which could be of the same order as $\Gamma(s)$ in general. However, only at one special point, i.e.~$\tanh\beta=1/d^2$, does the multi-region effect strictly vanishes $\Omega(\Delta)=0$, where \eqnref{eq: dSmin} reduces to
\eq{\label{eq: dSmin RUC}\partial_t S_\text{min}(|A|)=2\Gamma\big(\partial_{|A|}S_\text{min}(|A|)\big),}
which recovers the result in \refcite{Nahum2017Quantum,Jonay2018CDOSE}. This point corresponds to the random unitary circuit dynamics, where the entanglement dynamics admits the causal structure in \figref{fig: graph}(a), that the single-region entanglement is not causally affected by the multi-region entanglement, therefore one can arrive at a closed-form equation \eqnref{eq: dSmin RUC} within the single-region sector. Away from the $\tanh\beta=1/d^2$ point, \eqnref{eq: dSmin RUC} is incomplete, we will need to take into account the effect of multi-region entanglement as in \eqnref{eq: dSmin}.

To make broader connections, we study the entanglement velocity $v_\text{E}$, which is defined to be the growth rate of the entanglement entropy in half-infinite region on an open chain (with only one entanglement cut in the middle of the chain). It is therefore half of the entropy growth rate $\partial_t S_\text{min}$ in \eqnref{eq: dSmin}, i.e.~$v_\text{E}=\frac{1}{2}\partial_t S_\text{min}$. \refcite{Jonay2018CDOSE,Couch2019The-Speed} proposed the entanglement velocity $v_E(s)=\Gamma(s)$ as a function of the entropy gradient $s$. However, in more general cases, the entanglement velocity $v_\text{E}(s,\Delta)$ could also be affected by the multi-region entanglement, characterized by the entropy gap $\Delta$,
\eq{\label{eq: vE} v_\text{E}(s,\Delta)=\Gamma(s)+\Omega(\Delta).}

\begin{figure}[htbp]
\begin{center}
\includegraphics[width=0.85\columnwidth]{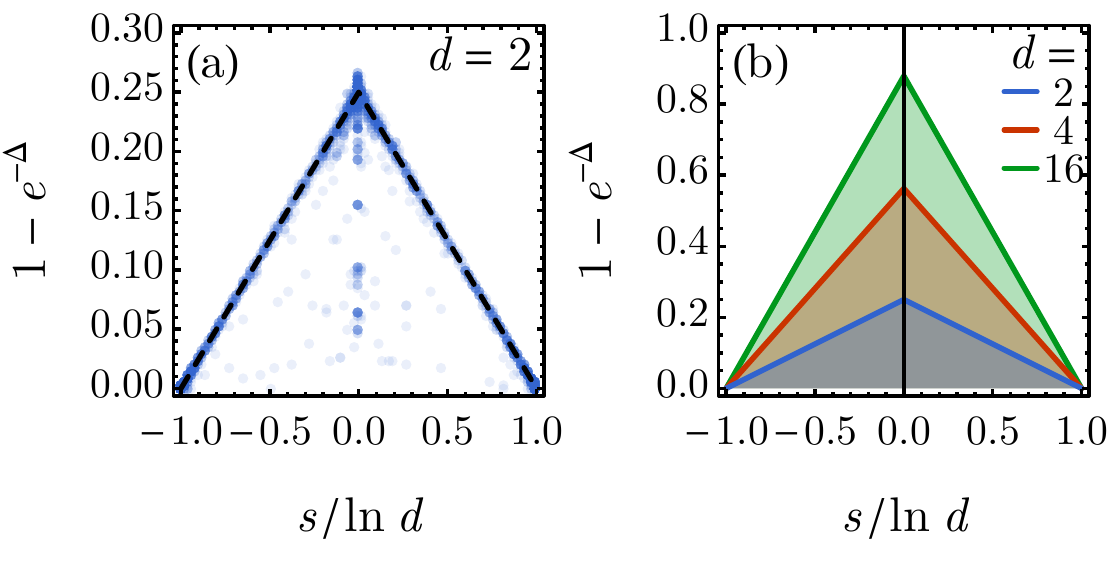}
\caption{(a) Collections of $s$ and $\Delta$ (in terms of $1-e^{-\Delta}$) under locally scrambled quantum dynamics at $d=2$. Data points mostly distributes within a triangle region. (b) The triangle region for different qudit dimensions $d$.}
\label{fig: pyramid}
\end{center}
\end{figure}

There is no definitive relation between the entropy gap $\Delta$ and the entropy gradient $s=\partial_{|A|}S_\text{min}$. However, as we collect data of $(s,\Delta)$ pairs over different region sizes $|A|$ at different times $t$ under different dynamics $\beta$ from different initial conditions, we find that they mostly lies in a triangle region, as depicted in \figref{fig: pyramid}(a) and described by
\eq{\label{eq: Omega bound}1-e^{-\Delta}\lesssim (1-1/d)^2(1-|s|/\ln d).} 
In fact, most of the data points concentrate along the triangle edges. A few outliers only appear at the tip of the triangle near $s=0$, which are generated in the early stage local thermalization (which will not show up in the coarse-grained long-time dynamics). The magnitude of the entropy gradient can be interpreted as the entropy density $|s|$, which can not exceed $\ln d$, as each qudit can at most contribute $\ln d$ entanglement entropy. The triangle region shape is only affected by the qudit dimension $d$ as shown in \figref{fig: pyramid}(b).

Given the result in \eqnref{eq: vE} and the bound in \eqnref{eq: Omega bound} and the definition , we can derive the following bound for the entanglement velocity
\eq{\label{eq: vE bound}v_\text{E}(s,\Delta)\leq\left\{
\begin{array}{ll}
\Gamma(s) & \tanh\beta\leq 1/d^2,\\
\Gamma(s)+\tilde{\Omega}(s) & \tanh\beta>1/d^2,
\end{array}\right.}
where $\tilde{\Omega}(s)$ is the upper-bound for $\Omega(\Delta)$ in terms of $s$,
\eq{\label{eq: tilde Omega}\tilde{\Omega}(s)=\frac{g(d-1)}{d^2(d+1)}(d^2 \sinh\beta-\cosh\beta)\Big(1-\frac{|s|}{\ln d}\Big).}
For $\tanh\beta\leq 1/d^2$, $\Omega(\Delta)$ is negative, meaning that the multi-region entanglement tends to slow down the entropy growth in this case, so $v_\text{E}$ is still bounded by $\Gamma(s)$ from above. For $\tanh\beta>1/d^2$, $\Omega(\Delta)$ is positive, meaning that the multi-region entanglement will speed up the entropy growth. Especially for large $\beta$, $\Omega(\Delta)$ can be much greater than $\Gamma(s)$ and dominates the contribution to $v_\text{E}$. In this case, we should bound $v_E$ by $\Gamma(s)+\tilde{\Omega}(s)$, where $\tilde{\Omega}(s)$ provides an upper bound for $\Omega(\Delta)$ according to the observation in \eqnref{eq: Omega bound}. In the following, we will further calculate the butterfly velocity $v_\text{B}$ and compare it with $v_\text{E}$ to examine the validity of the velocity inequality $v_\text{E}\leq(\ln d-|s|)v_\text{B}$ proposed in \refcite{Couch2019The-Speed}.

\section{Operator Dynamics in Locally Scrambled Quantum Systems}\label{operator-dynamics}

\subsection{Out-of-Time-Order Correlator}

The entanglement feature formalism not only describes the entanglement dynamics of quantum states, but also applies to the dynamics of operator spreading under locally scrambled quantum systems. Let $U(t)$ describe the unitary time evolution operator by time $t$. In the Heisenberg picture, a local Hermitian operator $O_i$ on site-$i$ will evolve as $O_i(t)=U(t)O_iU(t)^\dagger$. We are interested in the operator-averaged out-of-time-order correlator (OTOC) at infinite temperature
\eq{\label{eq: def OTOC}\OTOC(i,j;t)=\mathop{\dsE}\limits_{O_i,O_j}\Tr O_i(t)O_jO_i(t)O_j,}
which provides one way to quantify scrambling by probing how an operator $O_i(t)$ grows with time. For locally scrambled quantum dynamics, the operator $O_i(t)$ is expected to expand ballistically with a butterfly velocity $v_\text{B}$, which, in this case, is also the Lieb-Robinson velocity as the OTOC is calculated in the infinite temperature limit. The butterfly velocity $v_\text{B}$ can be extracted from the causal light-cone-like structure of the OTOC in the spacetime. We will calculate it using the entanglement feature formalism as follows.

The operator-averaged OTOC only depends on the operator entanglement of the unitary evolution $U(t)$\cite{Li2017MOCNMRQS,Lensky2018Chaos}, which can be captured by the entanglement feature operator $W_{U(t)}$, defined as
\eq{W_{U(t)}=\sum_{A,A'}\ket{A}e^{-S_{U(t)}(A,A')}\bra{A'},}
where $S_{U(t)}(A,A')$ denotes the entanglement entropy of the operator $U(t)$ (under the operator-state mapping)\cite{Prosen2007Operator,Hosur2016Chaos,Zhou2017Operator,Nie2018Signature} over region $A'$ on the input (past) side and region $A$ on the output (future) side. As derived in  
\refcite{You2018Entanglement}, the OTOC in \eqnref{eq: def OTOC} can be obtained from the entanglement feature operator $W_{U(t)}$ as
\eq{\label{eq: OTOC=W}\OTOC(i,j;t)=d^{-(N+2)}\bra{i}W_{U(t)}P\ket{j},}
where $\ket{i}$ denotes the Ising basis state $\ket{[\sigma]}$ with a down-spin at site-$i$ and up-spin elsewhere, i.e.~$\sigma_i=-1$ and $\sigma_j=+1$ for $j\neq i$, and $P=\prod_i X_i$ is the global spin flip operator.

Now we restrict $U(t)$ to the locally scrambled quantum dynamics, whose corresponding entanglement dynamics is described by the entanglement feature Hamiltonian $H_\text{EF}$, then according to \refcite{Kuo2019Markovian}, the entanglement feature operator $W_{U(t)}$ will be given by
\eq{W_{U(t)}W_\id^{-1}=e^{-tH_\text{EF}}.}
So the OTOC in \eqnref{eq: OTOC=W} can be expanded as
\eqs{\label{eq: OTOC expand}
\OTOC(i,j;t)&=d^{-(N+2)}\bra{i}e^{-t H_\text{EF}}W_\id P\ket{j}\\
&=\sum_{k=0}^{\infty}\frac{(-t)^k}{k! d^{N+2}}\bra{i} H_\text{EF}^kW_\id P\ket{j},}
where $W_\id=\prod_i(d^2+d X_i)$ is the entanglement feature operator for the identity operator. For convenience, we define $x\equiv|i-j|$ to be the distance between $i$ and $j$ sites, and the OTOC will only depend on $x$ (and $t$) given the translation symmetry of $H_\text{EF}$. For the $k=0$ term, $\bra{i}W_\id P\ket{j}=d^{N+2}$ naturally cancels the denominator. For $k>0$ terms, the first non-vanishing contribution comes at the $k=x$ order, because it takes at least $x$ steps of local operations in \figref{fig: dynamics} to transform $\ket{j}$ to $\ket{i}$ (by moving the entanglement cuts all the way from $j$ to $i$). Careful analysis shows that to the leading order in time, we have
\eq{\label{eq: OTOC}\OTOC(x,t)=1-(1-d^{-2})\frac{(tg\cosh\beta)^x}{x!}+\scO(t^{x+1}).}
Its detailed derivation can be found in \appref{app: OTOC}.

\subsection{Butterfly Velocity and Velocity Inequality}

To extract the butterfly velocity $v_B$, we examine the velocity-dependent OTOC\cite{Khemani2018Velocity-dependent} by setting $x=v t$ in $\OTOC(x,t)$. If $v$  happens to match the butterfly velocity, the OTOC will remain constant along the velocity cut in the long time limit (as the OTOC is riding on the kink front of the light-cone). According to \eqnref{eq: OTOC},
\eq{\lim_{t\to\infty}\OTOC(vt,t)=\lim_{t\to\infty}1-(1-d^{-2})\Big(\frac{g\cosh\beta}{v}\Big)^{vt},}
such that a finite limit of $\lim_{t\to\infty}\OTOC(vt,t)$ (i.e.~neither vanishing or diverging) is achieved when and only when $(g/v)\cosh\beta=1$, thus the butterfly velocity reads
\eq{\label{eq: vB}v_\text{B}=g\cosh\beta.}

\begin{figure}[htbp]
\begin{center}
\includegraphics[width=0.78\columnwidth]{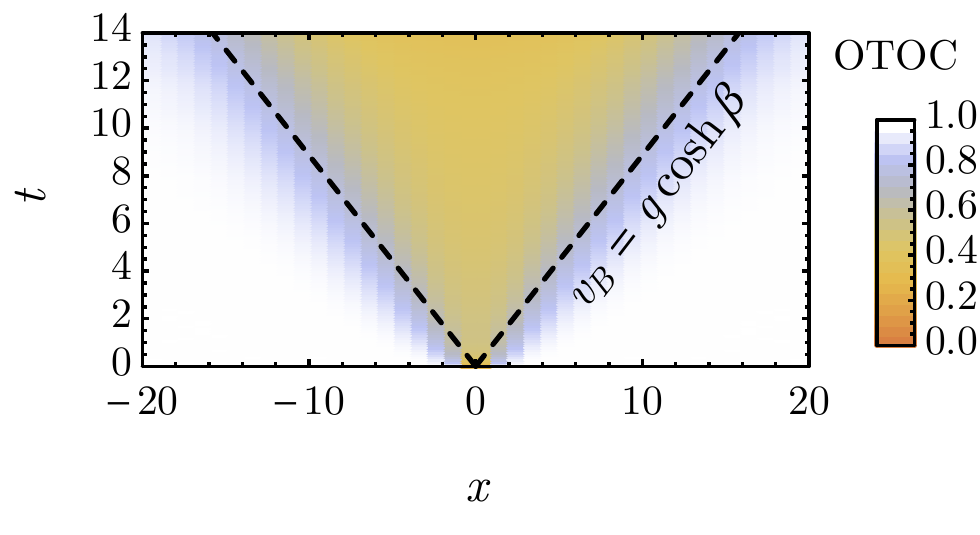}
\caption{OTOC for locally scrambled quantum dynamics at $g=1$ and $\beta=0.5$ on a 100-site lattice, calculated by the MPS-based numerical approach with a MPS bond dimension $D=192$.}
\label{fig: OTOC}
\end{center}
\end{figure}

We can verify this result by MPS-base numerical evaluation of \eqnref{eq: OTOC=W}. We start with the MPS representation of the initial state $W_\id P\ket{j}$ and apply the evolution operator $e^{-t H_\text{EF}}=\prod e^{-\delta t H_\text{EF}}$ in its Trotterized form. Following the time evolving block decimation (TEBD)\cite{Vidal2004Efficient,Zwolak2004Mixed-State,Verstraete2004Matrix} algorithm, we update the MPS tensors by truncated singular value decomposition method. The final MPS state is then overlapped with the $\bra{i}$ state to extract the OTOC. The result is shown in \figref{fig: OTOC}, which confirms the expression \eqnref{eq: vB} of the butterfly velocity $v_B$.

We can compare the butterfly velocity $v_B$ and the entanglement velocity $v_E(s,\Delta)$ to test the following velocity inequality
\eq{\label{eq: vEvB} v_E(s,\Delta)\leq (\ln d-|s|)v_B,}
which was originally proposed in \refcite{Couch2019The-Speed} in the context of AdS/CFT and in \refcite{Mezei2018Membrane,Jonay2018CDOSE} for the membrane models of entanglement growth. Based on our previous discussion, we have known that the entanglement velocity not only depends on single-region entanglement features like the entropy gradient $s=\partial_{|A|}S_\text{min}$, but also depends on multi-region entanglement features like the entropy gap $\Delta$ defined in \eqnref{eq: Delta}. We would like to check if $v_E(s,\Delta)$ can be universally bounded by the $\Delta$-independent right-hand-side $(\ln d-|s|)v_B$ of \eqnref{eq: vEvB}.

\begin{figure}[htbp]
\begin{center}
\includegraphics[width=\columnwidth]{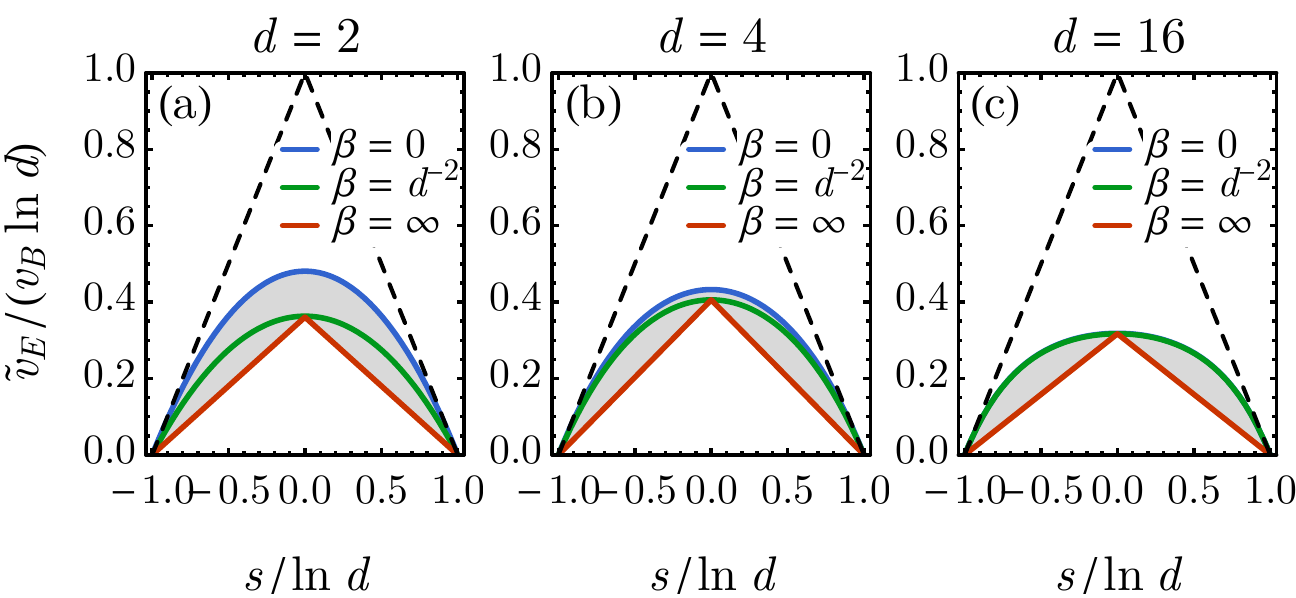}
\caption{The velocity ratio $\tilde{v}_E/(v_B \ln d)$ defined in \eqnref{eq: ratio ineq} v.s. the entropy gradient $s$ for different $\beta$ parameter and different qudit dimensions (a) $d=2$, (b) $d=4$, (c) $d=16$. The dashed line indicates the upper-bound $(1-|s|/\ln d)$. The gray area is the area swept by the curve as $\beta$ varies from $0$ (blue) to $\infty$ (red). }
\label{fig: velocity}
\end{center}
\end{figure}

Given the entanglement velocity bound in \eqnref{eq: vE bound}, we only need to check a more restrictive inequality
\eq{\label{eq: ratio ineq}\frac{\tilde{v}_E(s)}{v_B\ln d}\equiv\frac{\Gamma(s)+\tilde{\Omega}(s)}{v_B\ln d}\leq 1-\frac{|s|}{\ln d},}
where $\Gamma(s)$ is given in \eqnref{eq: dSmin} and $\tilde{\Omega}(s)$ is given in \eqnref{eq: tilde Omega}. It is understood that for $\tanh\beta\leq 1/d^2$, the $\tilde{\Omega}(s)$ term is automatically switched off. $\tilde{v}_E(s)=\Gamma(s)+\tilde{\Omega}(s)$ provides a $\Delta$-independent upper-bound for $v_E(s,\Delta)$, which effectively maxing out the multi-region entanglement effect. If the velocity ratio $\tilde{v}_E/(v_B\ln d)$ satisfies the inequality \eqnref{eq: ratio ineq},  the velocity inequality in \eqnref{eq: vEvB} will also hold. As we show in \figref{fig: velocity}, this is indeed the case for all qudit dimensions $d$ at all levels of entropy density $|s|$, under the locally scrambled quantum dynamics of any $g$ and $\beta$ parameters.

To gain a better analytic understanding, we notice that the velocity inequality is tight when $|s|\to\ln d$ and  $\beta\to 0$, where the bounding line is tangent to the velocity ratio curve in \figref{fig: velocity}. Given that $\tilde{v}_E(s)$ is a concave function of $s$, it is actually sufficient to check that its (negative) slope $-\partial_s \tilde{v}_E$ at the $s=\ln d$ corner is smaller than $v_B$, which allows us to obtain some simpler analytic results. We can show that
\eq{-\partial_s\tilde{v}_\text{E}|_{s=\ln d}\leq\left\{
\begin{array}{ll}
ge^{-\beta} & \tanh\beta\leq 1/d^2,\\
ge^{-\beta}+\tilde{\Omega}' & \tanh\beta>1/d^2,
\end{array}\right.}
where the derivative $\tilde{\Omega}'$ is given by
\eq{\tilde{\Omega}'=\frac{(d-1)g\cosh\beta}{(d+1)\ln d}(\tanh\beta-1/d^2).}
For $\tanh\beta\leq 1/d^2$, it is obvious that $-\partial_s\tilde{v}_\text{E}|_{s=\ln d}\leq ge^{-\beta} \leq g \cosh\beta=v_B$, meaning that the slope is within the bound. For $\tanh\beta> 1/d^2$, we have
\eqs{\label{eq: slope1}-\partial_s\tilde{v}_\text{E}|_{s=\ln d}&\leq ge^{-\beta}+\tilde{\Omega}'\\
&=g\cosh\beta\Big(1-\frac{\alpha}{d^{2}}-(1-\alpha)\tanh\beta\Big),}
where $\alpha=(d-1)/((d+1)\ln d)$ is a $d$-dependent constant satisfying $0<\alpha<1$. Given that $\tanh\beta> 1/d^2$ in this case, \eqnref{eq: slope1} can be relaxed to
\eqs{\label{eq: slope2}-\partial_s\tilde{v}_\text{E}|_{s=\ln d}&<g\cosh\beta\Big(1-\frac{\alpha}{d^{2}}-\frac{1-\alpha}{d^2}\Big)\\
&=g\cosh\beta(1-d^{-2}),}
which is still smaller than $v_B=g\cosh\beta$. Thus we have proven \eqnref{eq: ratio ineq} and hence \eqnref{eq: vEvB} follows. In conclusion, our result shows that the velocity inequality \eqnref{eq: vEvB} holds for all locally scrambled quantum dynamics (including quantum Brownian dynamics and random unitary circuits).

\section{Summary}

In this work, we study the evolution of bipartite entanglement entropy under locally scrambled quantum dynamics. We point out the importance of the multi-region entanglement in describing the entanglement dynamics. We show that the effect of multi-region entanglement can modify or even dominate the entanglement growth. The common assumption that the entanglement growth rate is only a function of the local entanglement entropy gradient is shown to be incomplete, as it ignores the multi-region entanglement. We identify the explicit contribution from the multi-region entanglement to the entanglement dynamics. We show that the entanglement feature approach reduces to the entanglement membrane approach if the multi-region entanglement ignored, which clarifies the difference and relations between these two approaches.

Our systematic characterization of the multi-region entanglement is based on the recent development of the entanglement feature formalism, which organize the entanglement entropies over all possible bipartitions into a many-body state. We further notice that such a many-body state can be efficiently represented by matrix product states, which could enable efficient numerical simulation of the entanglement dynamics. We propose a two-parameter matrix product state ansatz to capture all the multi-region entanglement entropy. We provide physical interpretations of the ansatz parameters and study their evolution under locally scrambled quantum dynamics. We show that evolution generally consists of an early-stage local thermalization and a late-stage global thermalization with distinct dynamics signatures.

We also gain a deeper understanding of the physical meaning of the parameters in the entanglement feature Hamiltonian by classifying and comparing the causal structure of the entanglement dynamics. Our analysis indicates that different values of the parameter $\beta$ in the entanglement feature Hamiltonian could correspond to different types of quantum dynamics. Thus different models in the field of non-equilibrium quantum dynamics, such as the quantum Brownian dynamics and the random unitary circuits, are unified within the scope of locally scrambled quantum dynamics, and can be discussed in a more systematic manner.
 
We calculate the operator-averaged out-of-time-order correlator in the infinite temperature limit for locally scrambled quantum dynamics, from which we extract the butterfly velocity and establish its dependence on the $\beta$ parameter. Despite of the multi-region region entanglement effect, we still find that the previously conjectured inequality between the entanglement velocity and the butterfly velocity remains valid for all values of $\beta$.

For future works, it is desired to extend the current approach to more general dynamics beyond the locally scrambled quantum dynamics, or to incorporate symmetry into entanglement dynamics. It will also be interesting to explore the entanglement dynamics in higher dimensions, where the entanglement feature state could exhibit topological order.

\begin{acknowledgements}
We acknowledge the discussions with Xiao-Liang Qi, John McGreevy, and Tsung-Cheng Lu. We also thank Wei-ting Kuo, Dan Arovas for collaborating on related subjects in an earlier work. AAA and YZY are supported by a startup fund from UCSD.
\end{acknowledgements}
\bibliographystyle{apsrev4-1} 
\bibliography{ref}
\onecolumngrid

\newpage
\appendix
\section{Efficacy of the $D=2$ MPS}\label{app: D=2}

\begin{figure}[htbp]
	\begin{center}
		\includegraphics[height=0.17\textheight]{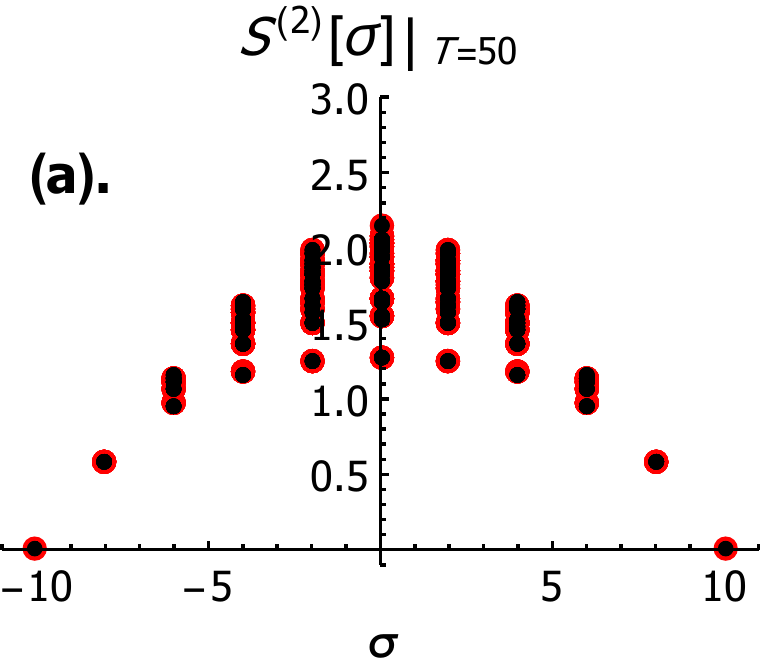}\qquad
		\includegraphics[height=0.17\textheight]{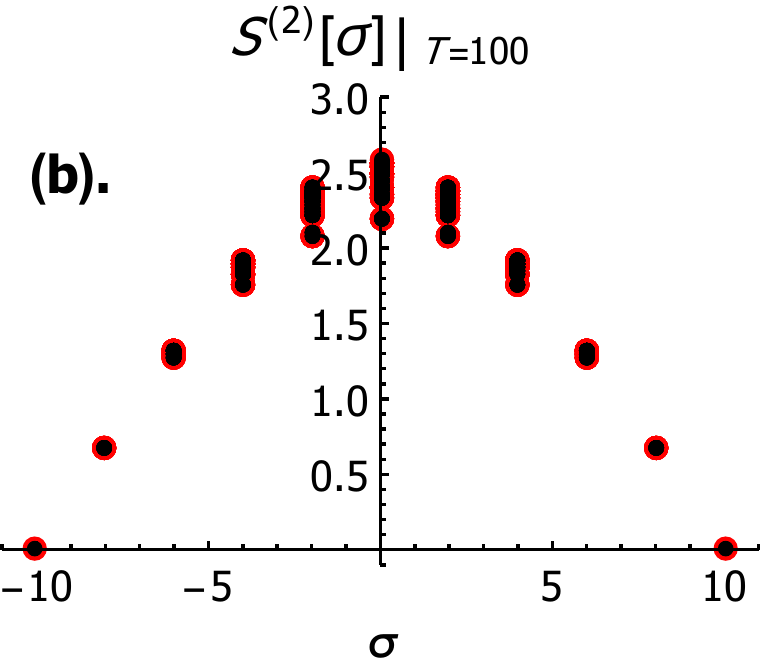}\qquad
		\includegraphics[height=0.17\textheight]{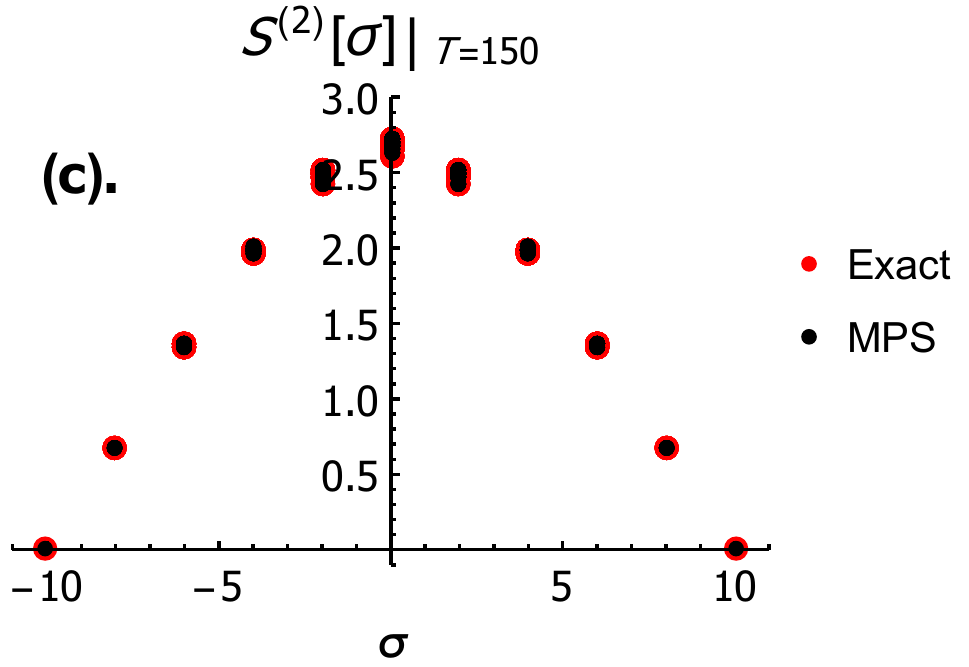}
		\caption{The three figures show the evolution of the Renyi entropy in the fractional swap gate for swap probability $x=0.1$ at three different time slices: $T=50,100$ and $150$ for figures (a),(b), and (c) respectively. The $x$-axis is the magnetization, which is related to the size of the entangling region by $|A|=(\sigma+N)/2$. Initially, at $T=0$, the state starts off as a product state and all bipartite entropies are zero. Then, as it evolves, it takes the generic form of the multi-region entanglement continuum, lower-bounded by the single-region entanglement which is area-law. This can be observed in (a). It gradually becomes dominated by the volume law term, at which point the area law plateau begins to vanish, as seen in (b). As it approaches the Page state, which is a pure volume law state, the multi-region entanglement continuum collapses into a single curve so that $S^{(2)}(A)=S^{(2)}_{\text{min}}(|A|)$, as we can see in (c). The red points are given by the exact numerics, whereas the black points are given by the $D=2$ MPS evolved by TEBD. We see nearly perfect alignment between the two for the full multi-region entanglement continuum for arbitrary choice of EF parameters $g,\beta$.}
		\label{fig: performance}
	\end{center}
\end{figure}

We can use the TEBD approach to evolve a $D=2$ MPS state to calculate the entanglement dynamics governed by $-\partial_t\ket{W_\Psi}=H_\text{EF}\ket{W_\Psi}$. By comparing MPS the result with the exact numerical solution the differential equation, we found that $D=2$ successfully captures the evolution of all the multi-region entanglement over the entire process of thermalization from a product state, as shown in \figref{fig: performance}. 

We find that the $D=2$ MPS captures the full entanglement continuum regardless of the parameters $g,\beta$ of locally scrambled evolution. This is expected from the sign structure of the entanglement feature states\cite{Grover2015Entanglement}. Since generic positive vectors have constant law entanglement, it is expected that generic states (states that are not fine-tuned) will be well described by MPS. The exception is near the entanglement transition where the bond dimension is expected to diverge \cite{Fan2020Self-Organized}.

\begin{figure}[htbp]
	\begin{center}
		\includegraphics[width=0.6\textwidth]{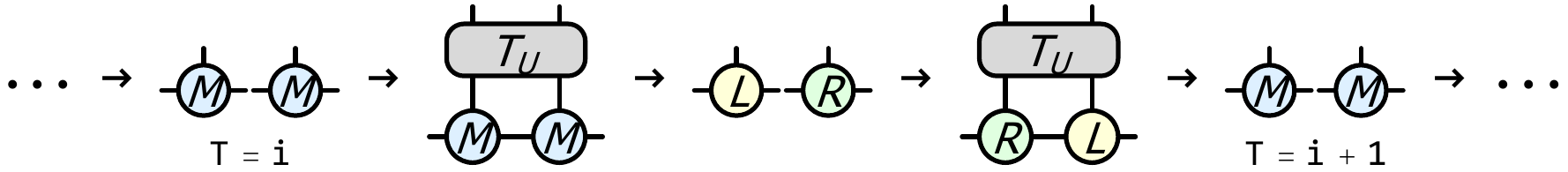}
		\caption{The Time Evolved Block Decimation (TEBD) procedure for updating the tensor $M$ in the translationally invariant MPS EF state. First, we apply an even layer of transfer matrices on the current state which is given by a product of $M$ tensors. After SVD truncation on the resulting tensor, we procure two new tensors $L$ and $R$. This is by taking the left (right) legs as the input (output) legs of a matrix and calculating its SVD decomposition $usv^{\dagger}$, truncating all but the two largest singular values ($D=2$), separating the result into a product of two matrices $u\sqrt{s}$ and $\sqrt{s}v^{\dagger}$, and then reshaping these resulting matrices into $L$ and $R$, respectively. Note that these tensors may break the translational symmetry in the system. Applying the odd layer now, the right tensor of the previous layer is the left tensor of the new layer. We now do SVD truncation once more to get the new $M$ tensors. These resulting tensors should be the same on the left and right, and so it suffices to just take the left or the right one. In principle, because the circuit breaks the translational symmetry into a two-site translational symmetry, it is possible that the tensors are not the same. However, we find that translational symmetry in the MPS is not broken as we evolve the circuit.}
		\label{fig: TEBD}
	\end{center}
\end{figure}

TEBD on the MPS ansatz works as in \figref{fig: TEBD}. We apply the two-local EF transfer matrix to the product of local MPS tensors. We separate the result into two tensors, $L$ and $R$ which are given by doing an SVD truncation on the product, taking only the two largest singular values. These tensors are then updated again, but in reversed order, by the transfer matrix according to the brick wall arrangement of the circuit. The resulting two tensors are identical and become the next MPS tensor $M$. 

\section{Derivation of Edges of Multi-Region Continuum}\label{app: edge}
We provide the detailed derivation of $S_\text{min}(|A|)$ and $S_\text{max}(|A|)$ here. Our starting point is the MPS ansatz $M_{(\alpha,\theta)}^\sigma$ in \eqnref{eq: M}, which can be written as
\eq{M^\sigma=e^{\alpha(\sin\theta\;X+\sigma\cos\theta\;Z)}.}
We have suppressed the subscript $(\alpha,\theta)$ for simplicity. We first evaluate the following matrix product
\eqs{(M^{\downarrow})^m (M^{\uparrow})^n&=e^{\alpha m(\sin\theta\;X-\cos\theta\;Z)}e^{\alpha n(\sin\theta\;X+\cos\theta\;Z)}\\
&=\big(\cosh\alpha m+\sinh\alpha m(\sin\theta\;X-\cos\theta\;Z)\big)\big(\cosh\alpha n+\sinh\alpha n(\sin\theta\;X+\cos\theta\;Z)\big)\\
&=\cosh\alpha m\cosh\alpha n\big(1+\tanh\alpha m(\sin\theta\;X-\cos\theta\;Z)\big)\big(1+\tanh\alpha n(\sin\theta\;X+\cos\theta\;Z)\big)\\
&=\cosh\alpha m\cosh\alpha n\big(1+(\tanh\alpha m+\tanh\alpha n)\sin\theta\;X+(\tanh\alpha m-\tanh\alpha n)\cos\theta\;Z\\
&\phantom{=\cosh\alpha m\cosh\alpha n\big(}\tanh\alpha m\tanh\alpha n(-\cos2\theta-\ii\sin2\theta\;Y)\big)\\
&=\cosh\alpha m\cosh\alpha n(c_0+c_1 X+\ii c_2 Y+ c_3 Z),}
where we have introduced the coefficients $c_{0,1,2,3}$ as
\eqs{c_0&=1-\tanh\alpha m\tanh\alpha n\cos2\theta,\\
c_1&=(\tanh\alpha m+\tanh\alpha n)\sin\theta,\\
c_2&=-\tanh\alpha m\tanh\alpha n\sin2\theta,\\
c_3&=(\tanh\alpha m-\tanh\alpha n)\cos\theta.\\}
The eigenvalues of $(M^{\downarrow})^m (M^{\uparrow})^n$ are given by $\mu_{\pm}=c_{\pm}\cosh\alpha m\cosh\alpha n$ with
\eq{c_\pm=c_0\pm\sqrt{c_1^2-c_2^2+c_3^2}.}
We notice that
\eqs{c_1^2-c_2^2+c_3^2&=(\tanh\alpha m+\tanh\alpha n)^2\sin^2\theta+(\tanh\alpha m-\tanh\alpha n)^2\cos^2\theta-\tanh^2\alpha m\tanh^2\alpha n\sin^2 2\theta\\
&=(\tanh^2\alpha m+\tanh^2\alpha n)-2 \tanh\alpha m \tanh\alpha n\cos 2\theta-\tanh^2\alpha m\tanh^2\alpha n\sin^2 2\theta\\
&=(1-\tanh\alpha m \tanh\alpha n\cos 2\theta)^2-(1-\tanh^2\alpha m)(1-\tanh^2\alpha n)\\
&=c_0^2-\sech^2\alpha m \sech^2\alpha n.}
So we have the following relation
\eq{\sqrt{{c_+}{c_-}}=\sqrt{c_0^2-(c_1^2-c_2^2+c_3^2)}=\sech\alpha m \sech\alpha n.}
These results are useful to evaluate the trace of $((M^{\downarrow})^m (M^{\uparrow})^n)^p$. Given the eigenvalues $\mu_{\pm}$ of $(M^{\downarrow})^m (M^{\uparrow})^n$,
\eqs{\label{eq: Tr1}\Tr ((M^{\downarrow})^m (M^{\uparrow})^n)^p&=\mu_+^p+\mu_-^p\\
&=\cosh^p\alpha m\cosh^p\alpha n(c_+^p+c_-^p)\\
&=\cosh^p\alpha m\cosh^p\alpha n(\sqrt{{c_+}{c_-}})^p((\tfrac{c_+}{c_-})^{p/2}+(\tfrac{c_-}{c_+})^{p/2})\\
&=\cosh^p\alpha m\cosh^p\alpha n \sech^p\alpha m \sech^p\alpha n((\tfrac{c_+}{c_-})^{p/2}+(\tfrac{c_-}{c_+})^{p/2})\\
&=(\tfrac{c_+}{c_-})^{p/2}+(\tfrac{c_-}{c_+})^{p/2}.}
If we define a new parameter $\eta$ via $e^\eta=(\tfrac{c_+}{c_-})^{1/2}$, the trace in \eqnref{eq: Tr1} can be written as
$\Tr ((M^{\downarrow})^m (M^{\uparrow})^n)^p=e^{\eta p}+e^{-\eta p}=2\cosh\eta p$. In particular,
\eqs{\cosh\eta&=\tfrac{1}{2}\big((\tfrac{c_+}{c_-})^{1/2}+(\tfrac{c_-}{c_+})^{1/2}\big)\\
&=\frac{{c_+}+{c_-}}{2\sqrt{{c_+}{c_-}}}\\
&=c_0\cosh\alpha m\cosh\alpha n\\
&=\cosh\alpha m\cosh\alpha n-\sinh\alpha m\sinh\alpha n\cos2\theta\\
&=(\cosh\alpha m\cosh\alpha n+\sinh\alpha m\sinh\alpha n)\sin^2\theta+(\cosh\alpha m\cosh\alpha n-\sinh\alpha m\sinh\alpha n)\cos^2\theta\\
&=\sin^2\theta\cosh\alpha (m+n)+\cos^2\theta\cosh\alpha (m-n).}
In conclusion, we have arrived at a trace formula
\eqs{&\Tr ((M^{\downarrow})^m (M^{\uparrow})^n)^p=2\cosh\eta p,\\
&\text{with }\cosh\eta=\sin^2\theta\cosh\alpha (m+n)+\cos^2\theta\cosh\alpha (m-n).}

Using the trace formula, we can now evaluate the bottom and upper edge entanglement entropies based on the MPS ansatz. We first calculate $S_\text{min}(|A|)$,
\eqs{S_\text{min}(|A|)&=-\ln\Tr(M^{\downarrow})^{|A|}(M^{\uparrow})^{N-|A|}+S_0\\
&=-\ln\big(2(\sin^2\theta\cosh\alpha N+\cos^2\theta\cosh\alpha(N-2|A|))\big)+\ln(2\cosh\alpha N)\\
&=-\ln\big(\sin^2\theta+\cos^2\theta\tfrac{\cosh\alpha(N-2|A|)}{\cosh\alpha N}\big),}
which gives \eqnref{eq: Smin}. We then calculate $S_\text{max}(|A|)$,
\eqs{S_\text{max}(|A|)&= -\ln\Tr\big(M^{\downarrow}(M^{\uparrow})^{N/|A|-1}\big)^{|A|}+S_0\\
&=-\ln(2\cosh\eta|A|)+\ln(2\cosh\alpha N)\\
&=-\ln\frac{\cosh \eta |A|}{\cosh\alpha N},}
where $\eta$ is set by $\cosh\eta=\sin^2\theta\cosh\alpha\tfrac{N}{|A|}+\cos^2\theta\cosh\alpha\big(\tfrac{N}{|A|}-2\big)$, as claimed in \eqnref{eq: Smax}.

Now we consider the thermodynamic limit ($N\to+\infty$) of $S_\text{min}(|A|)$ and $S_\text{max}(|A|)$. We note that
\eqs{\frac{\cosh\alpha(N-2|A|)}{\cosh\alpha N}&=\frac{\cosh\alpha N\cosh 2\alpha|A|-\sinh\alpha N\sinh 2\alpha|A|}{\cosh\alpha N}\\
&=\cosh 2\alpha|A|-\tanh\alpha N\sinh 2\alpha|A|\\
&\xrightarrow{N\to+\infty}\cosh 2\alpha|A|-\sinh 2\alpha|A|
=e^{-2\alpha|A|},}
therefore $S_\text{min}(|A|)=-\ln(\sin^2\theta+\cos^2\theta \;e^{-2\alpha|A|})$ as in \eqnref{eq: S largeN}. Similarly we have
\eqs{\cosh\eta&=\Big(\sin^2\theta+\cos^2\theta\frac{\cosh\big(\tfrac{\alpha N}{|A|}-2\alpha\big)}{\cosh\tfrac{\alpha N}{|A|}}\Big)\cosh\frac{\alpha N}{|A|}\\
&\xrightarrow{N\to+\infty}(\sin^2\theta+\cos^2\theta\;e^{-2\alpha})\cosh\frac{\alpha N}{|A|}.}
Take the inverse $\cosh$ function on both sides,
\eqs{\eta&=\arccosh\Big((\sin^2\theta+\cos^2\theta\;e^{-2\alpha})\cosh\frac{\alpha N}{|A|}\Big)\\
&\xrightarrow{N\to+\infty}\frac{\alpha N}{|A|}+\ln(\sin^2\theta+\cos^2\theta\;e^{-2\alpha}),}
therefore
\eqs{S_\text{max}(|A|)&=-\ln\frac{\cosh \eta |A|}{\cosh\alpha N}\\
&=-\ln\frac{\cosh(\alpha N+|A|\ln(\sin^2\theta+\cos^2\theta\;e^{-2\alpha}))}{\cosh\alpha N}\\
&\xrightarrow{N\to+\infty}-\ln e^{|A|\ln(\sin^2\theta+\cos^2\theta\;e^{-2\alpha})}\\
&=-|A|\ln(\sin^2\theta+\cos^2\theta\;e^{-2\alpha}).}
as in \eqnref{eq: S largeN}.

\section{Dynamics of MPS Parameters}\label{app: MPS}

In this section, we will derive the dynamic equation for MPS parameters and explain the numerical details in solving the equation. Our starting point is the imaginary-time Schr\"odinger equation $-\partial_t\ket{W_\Psi}=H_\text{EF}\ket{W_\Psi}$, which governs the evolution of entanglement feature state $\ket{W_\Psi}$. As we represent $\ket{W_\Psi}$ as a $D=2$ MPS  proposed in \eqnref{eq: MPS}, we would like to approximate the time evolution generated by $H_\text{EF}$ without leaving the variational manifold of the MPS ansatz \eqnref{eq: M}. Let us denote the MPS parameters $(\alpha,\theta)$ jointly as a vector $q$. Within the variational manifold, the entanglement feature state could only evolve in the tangent plane as $\partial_t\ket{W_q}=\ket{\partial_aW_q}\dot{q}_a$, where $\dot{q}_a\equiv\partial_t q_a$ and $\partial_a\equiv\partial_{q_a}$. We seek the optimal choice of $\dot{q}_a$ such that $-\partial_t\ket{W_q}=-\ket{\partial_aW_q}\dot{q}_a$ best approximates $H_\text{EF}\ket{W_q}$. The solution is given by minimizing the loss function
\eq{\scL(\dot{q})=\|\ket{\partial_aW_q}\dot{q}_a+H_\text{EF}\ket{W_q}\|^2.}
To define the loss function $\scL$, we also need to specify how to take the norm of the entanglement feature state. It is desired that the inner product of entanglement feature states $(\bra{W_{q'}},\ket{W_q})$ is such defined that $H_\text{EF}$ is self-adjoint, i.e.~$(\bra{W_{q'}},H_\text{EF}\ket{W_q})=(\bra{W_{q'}}H_\text{EF}^\intercal,\ket{W_q})$. Since $H_\text{EF}\neq H_\text{EF}^\intercal$ itself is not transpose symmetric, the inner product must involve a non-trivial metric, which turns out to be given by the following operator
\eq{W_\id^{-1}=(\tanh\delta\sinh\delta)^Ne^{-\delta\sum_i X_i},}
where $\delta=\frac{1}{2}\ln\frac{d+1}{d-1}$ is fixed by the qudit dimension $d$ of the quantum system. Therefore the norm of $\ket{W_q}$ should be defined as $\|\ket{W_q}\|=\bra{W_q}W_\id^{-1}\ket{W_\Psi}$. So the loss function can be expanded as a quadratic form of $\dot{q}$,
\eq{\scL(\dot{q})=\dot{q}_a G_{ab}\dot{q}_b+2h_a\dot{q}_a+c,}
with the coefficients given by
\eqs{G_{ab}&=\bra{\partial_aW_q}W_\id^{-1}\ket{\partial_bW_q},\\
h_a&=\bra{\partial_aW_q}W_\id^{-1}H_\text{EF}\ket{W_q},\\
c&=\bra{W_q}H_\text{EF}^\intercal W_\id^{-1}H_\text{EF}\ket{W_q}.}
The minimum of $\scL(\dot{q})$ is determined by
\eq{\label{eq: Gq=h} G_{ab}\dot{q}_b=-h_a,}
which gives the dynamic equation in \eqnref{eq: MPS dyn}. It can be formally written as $\dot{q}=-G^{-1}h$, but $G$ can become singular in the thermodynamic limit $N\to\infty$, which requires more detailed treatment.

Now we explain how to evaluate $G$ and $h$ given the MPS parameters. Given the MPS $\ket{W_q}$
\eq{\ket{W_q}=\dia{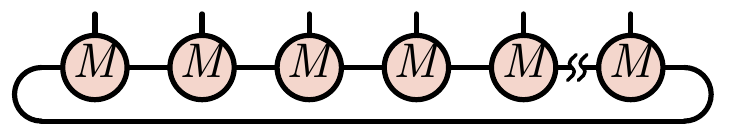}{27}{-12},}
$G$ and $h$ can be represented as the following tensor networks,
\eqs{G_{ab}&=\sum_{ij}\dia{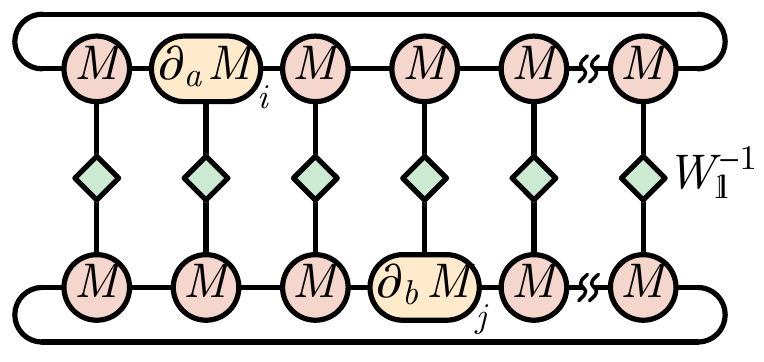}{68}{-32},\\
h_{a}&=\sum_{i}\sum_{\langle jk\rangle}\dia{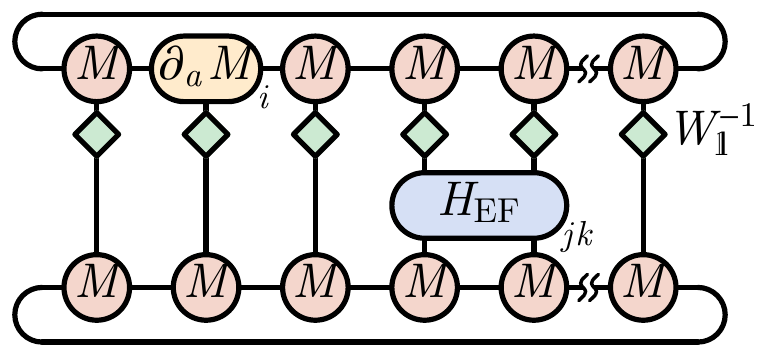}{68}{-32},}
where $M$ denotes the MPS tensor given in \eqnref{eq: M} and $(H_\text{EF})_{jk}$ denotes the term in $H_\text{EF}$ on the $\langle jk\rangle$ link which reads $\frac{g}{2}(1-Z_jZ_k)e^{-\delta(X_j+X_k)-\beta X_j X_k}$. The derivatives of the MPS tensor $M$ are given by
\eqs{\partial_\alpha M^{\sigma}&=\sinh \alpha\; I+\cosh\alpha (\sin \theta\; X+\sigma \cos \theta\; Z),\\
\partial_\theta M^{\sigma}&=\cosh \alpha\; I+\sinh\alpha (\cos \theta\; X -\sigma \sin \theta\; Z).\\}

To evaluate these tensor networks, we introduce the transfer operator
\eq{T=\dia{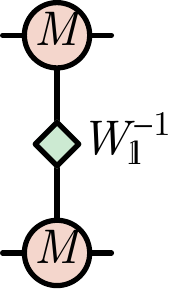}{56}{-27}.}
Let $\ket{\tau}$ be the leading eigenvector of $T$ with the eigenvalue 1 (if the leading eigenvalue of $T$ is not 1, we rescale $W_\id^{-1}$ to make it 1), such that
\eq{\dia{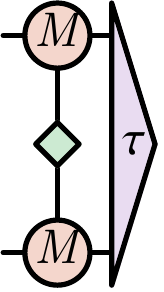}{56}{-27}=\dia{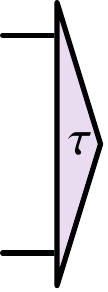}{56}{-27},\quad\dia{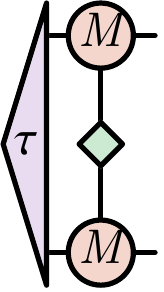}{56}{-27}=\dia{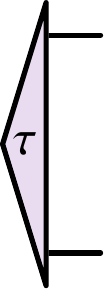}{56}{-27}.}
We denote the pseudo inverse of $(1-T)$ as $\Pi=(1-T)^{-1}$, such that in the large $N$ limit, the ladder diagram reads
\eq{\sum_{n=0}^{N}T^n=\dia{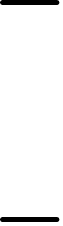}{43}{-20.5}+\dia{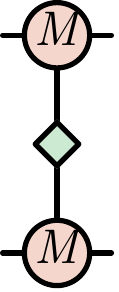}{56}{-27}+\dia{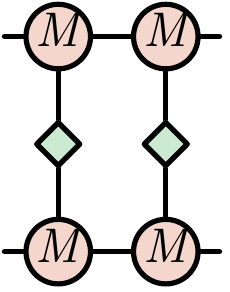}{56}{-27}+\dia{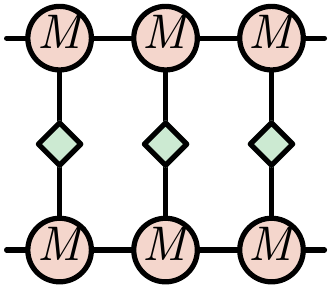}{56}{-27}+\cdots=\dia{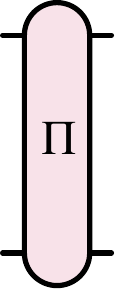}{56}{-27}+N\dia{tr}{56}{-27}\dia{tl}{56}{-27}.}
With these preparations, we can show that $G$ and $h$ scales with the system size $N$ in the following manner
\eqs{\label{eq: Gh in N}
G&=N G^{(1)}+N^2 G^{(2)},\\
h&=N h^{(1)}+N^2 h^{(2)},}
with $G^{(1,2)}$ and $h^{(1,2)}$ given by the following tensor networks
\eqs{\label{eq: Gh diags}G_{ab}^{(1)}&=\dia{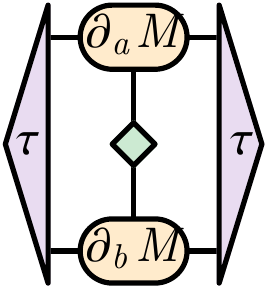}{56}{-27}+\dia{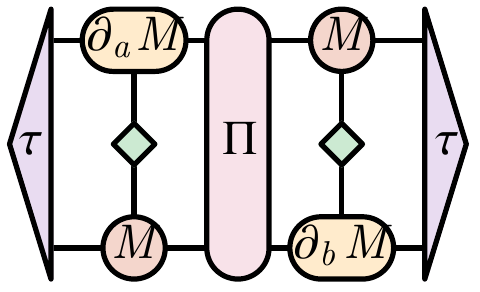}{56}{-27}+\dia{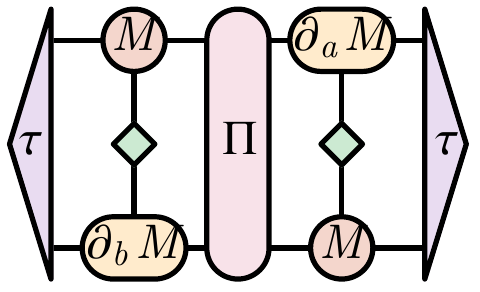}{56}{-27},\\
G_{ab}^{(2)}&=\dia{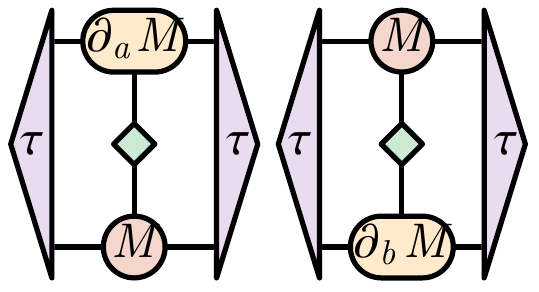}{56}{-27},\\
h_{a}^{(1)}&=\dia{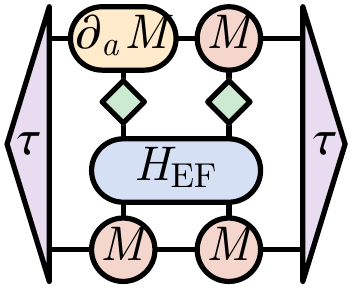}{56}{-27}+\dia{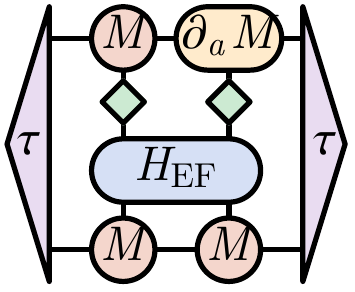}{56}{-27}+\dia{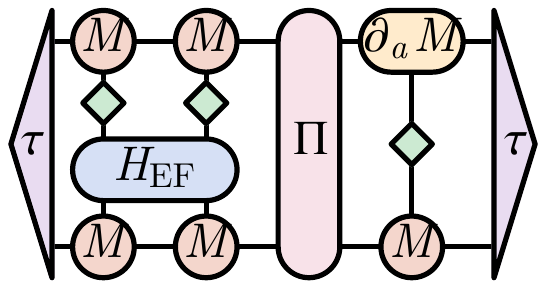}{56}{-27}+\dia{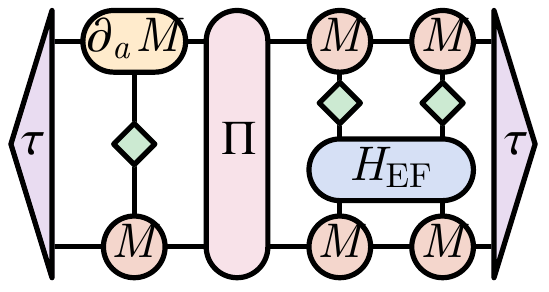}{56}{-27},\\
h_{a}^{(2)}&=\dia{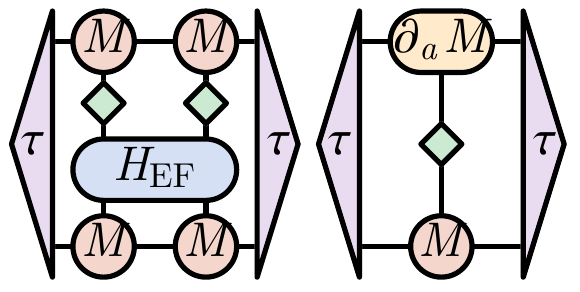}{56}{-27}.}
According to the dynamic equation \eqnref{eq: Gq=h}, time-derivatives of MPS parameters are determined by $\dot{q}=-G^{-1}h$. To calculate the inverse of $G$, we note that $G$ is a $2\times 2$ real symmetric matrix, so $G^{-1}$ can be expressed as
\eq{G^{-1}=\frac{2JGJ}{\Tr JGJG},}
where $J=\smat{0&1\\-1&0}$. Further using the form in \eqnref{eq: Gh in N}, we have
\eqs{\label{eq: q frac}\dot{q}&=-\frac{2JGJh}{\Tr JGJG}\\
&=-\frac{2\big(N^2 JG^{(1)}Jh^{(1)}+N^3 JG^{(2)}Jh^{(1)}+N^3 JG^{(1)}Jh^{(2)}+N^4 JG^{(2)}Jh^{(2)}\big)}{N^2 \Tr JG^{(1)}JG^{(1)}+2N^3 \Tr JG^{(1)}JG^{(2)}+N^4 \Tr JG^{(2)}JG^{(2)}}.}
An important observation is that $JG^{(2)}Jh^{(2)}=0$ and $JG^{(2)}JG^{(2)}=0$, because $G^{(2)}$ and $h^{(2)}$ has the structure of $G^{(2)}=\ket{\gamma}\bra{\gamma}$ and $h^{(2)}=\ket{\gamma}\eta$ where $\ket{\gamma}$ is a two-component vector and $\eta$ is a real number, such that the vanishing $\bra{\gamma}J\ket{\gamma}=0$ (due to the antisymmetric nature of $J$) results in the vanish of $JG^{(2)}Jh^{(2)}$ and $JG^{(2)}JG^{(2)}$. Then both the numerator and denominator of \eqnref{eq: q frac} is dominated by the $N^3$ term in the $N\to\infty$ limit. Therefore, we can evaluate the time derivative
$\dot{q}$ from
\eq{\dot{q}=-\frac{JG^{(2)}Jh^{(1)}+JG^{(1)}Jh^{(2)}}{\Tr JG^{(1)}JG^{(2)}}.}
By iteratively updating $q\to q+\dot{q}\,\dd t$ and calculating $\dot{q}$ from the diagrams in \eqnref{eq: Gh diags}, we can obtain the time-evolution of the MPS parameters numerically.

\section{Calculating OTOC}\label{app: OTOC}

Here we explain how the OTOC is calculated. We start from the expansion in \eqnref{eq: OTOC expand}
\eq{\label{eq: OTOC0}\OTOC(i,j;t)=\sum_{k=0}^{\infty}\frac{(-t)^k}{k! d^{N+2}}\bra{i} H_\text{EF}^kW_\id P\ket{j},}
where $W_\id=\prod_{i}(d^2+dX_i)$, $P=\prod_i X_i$, and $\ket{i}$ denotes the Ising basis state with a down-spin only at site-$i$ and up-spins elsewhere. The entanglement feature Hamiltonian takes the form of \eqnref{eq: HEF uvw},
\eq{H_\text{EF}=\sum_{\langle ij\rangle}\frac{1-Z_iZ_j}{2}(u-v(X_i+X_j)+w X_iX_j),}
with parameters $u,v,w$ given by \eqnref{eq: uvw},
\eq{\label{eq: uvw2}\left(\begin{array}{c}u\\v\\w\end{array}\right)=\frac{g\cosh\beta}{d^2-1}\left(\begin{array}{c}d^2-\tanh\beta\\d-d\tanh\beta\\1-d^2\tanh\beta\end{array}\right).}
We can combine the $d^{-N}$ factor and the operator $W_\id P$ together to define 
\eq{\label{eq: F}
F\equiv\frac{1}{d^N}W_\id P=\prod_i (1+ d X_i).}
Such that \eqnref{eq: OTOC0} can be simplified a little bit,
\eq{\label{eq: OTOC1}\OTOC(i,j;t)=\frac{1}{d^2}\sum_{k=0}^{\infty}\frac{(-t)^k}{k!}\bra{i} H_\text{EF}^kF\ket{j}.}
We will be able to evaluate $\bra{i} H_\text{EF}^kF\ket{j}$ for leading orders in $H_\text{EF}$, which will provide the OTOC in the short-time limit (the expansion can be thought as controlled by the small parameter $t$). The task to evaluate $\bra{i} H_\text{EF}^kF\ket{j}$ can be considered as how to connect the $\ket{j}$ state (a single down-spin at site-$j$) to the $\bra{i}$ state (a single down-spin at site-$i$) by applying the operator $F$ followed by a sequence of $H_\text{EF}$. The net effect is to move the down-spin from site-$j$ to site-$i$ on a background of all up-spins. 

\begin{figure}[htbp]
\begin{center}
\includegraphics[width=0.64\columnwidth]{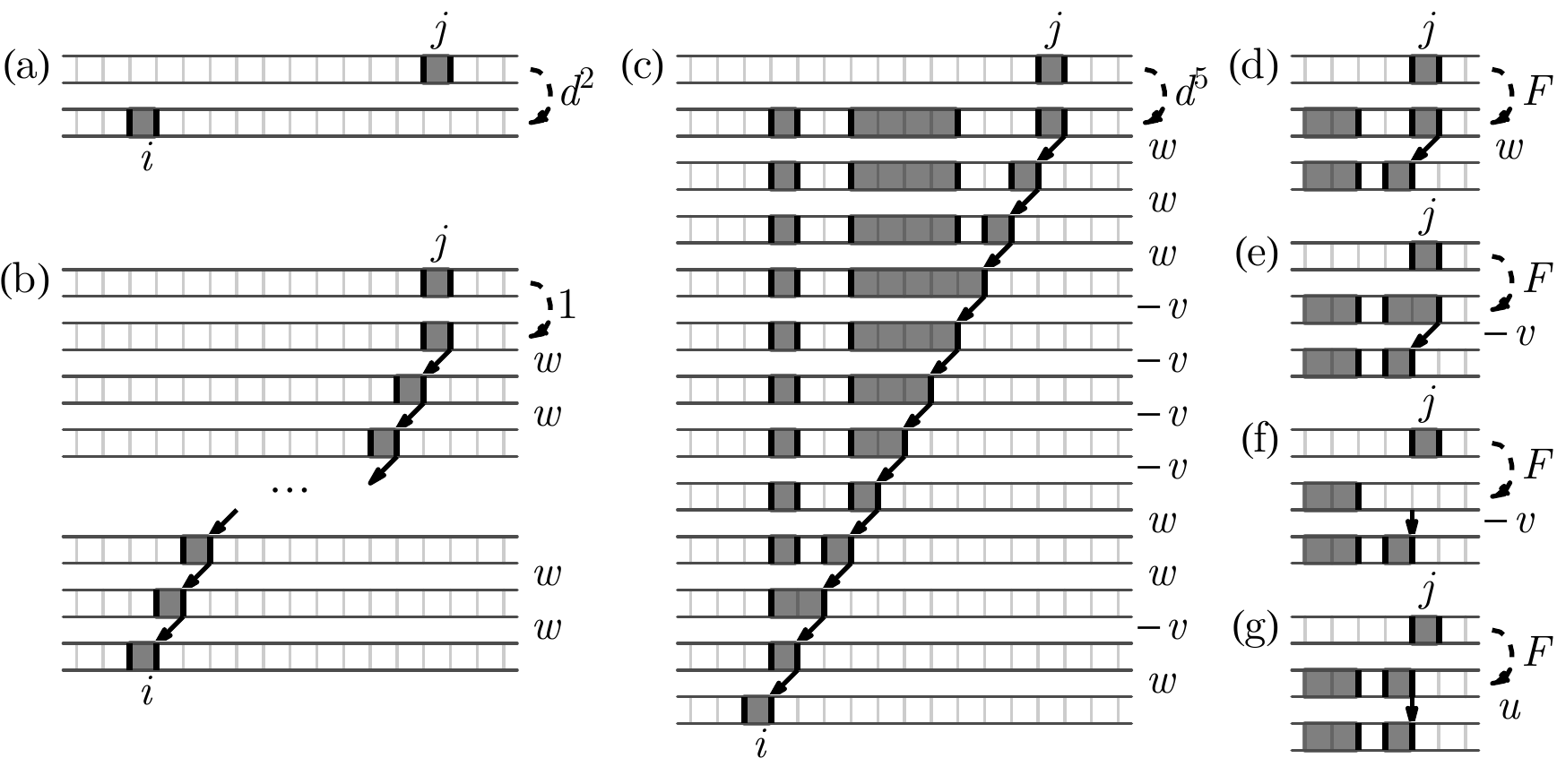}
\caption{Examples of entanglement region dynamics in calculating $\bra{i}H_\text{EF}^kF\ket{j}$.}
\label{fig: moves}
\end{center}
\end{figure}

To warm up, let us start with the 0th order term $\bra{i}F\ket{j}$ (i.e.~$k=0$). From \eqnref{eq: F}, we can see that $F$ is a non-local operator, which sprinkles spin flips with a coefficient $d$. It can be expanded as polynomials of $X_i$ operators. To connect $\ket{j}$ and $\ket{i}$ states, $F$ needs to remove the down-spin at site-$j$ and create the down-spin at site-$i$, as shown in \figref{fig: moves}(a). This corresponds to spin-flip operations at both sites, described by $X_iX_j$. The coefficient of the $X_iX_j$ term in the expansion of $F$ is $d^2$ (each $X$ operator contributes a factor $d$). So we have 
\eq{\label{eq: HF0}\bra{i}F\ket{j}=d^2.}

Suppose the sites $i$ and $j$ are spatially separated by the distance $x=|i-j|$, it turns out that the next leading contribution is at the order of $k=x$ as $\bra{i} H_\text{EF}^x F\ket{j}$, because all the lower order terms $\bra{i} H_\text{EF}^k F\ket{j}=0$ vanish for $0<k<x$. This has to do with the specific algebraic relation between $H_\text{EF}$ and $F$. As elaborated in \refcite{Kuo2019Markovian}, the entanglement feature Hamiltonian must satisfy the following defining properties: $H_\text{EF}W_\id=W_\id H_\text{EF}^\intercal$ and $H_\text{EF}P=PH_\text{EF}$. Based on the definition of $F$ in \eqnref{eq: F}, we have $H_\text{EF}F=F H_\text{EF}^\intercal$, therefore
\eq{\label{eq: HF=FH}\bra{i} H_\text{EF}^k F\ket{j}=\bra{i}  F {H_\text{EF}^{\intercal k}}\ket{j}.}
Since each term in $H_\text{EF}$ carries a projection operator $\frac{1-Z_iZ_j}{2}$, \eqnref{eq: HF=FH} implies that the left-most $H_\text{EF}$ must act on an entanglement cut in the $\bra{i}$ state and the right-most $H_\text{EF}$ must act on an entanglement cut in the $\ket{j}$ state. Similar restrictions applies to all the intermediate actions of $H_\text{EF}$. On one hand, $H_\text{EF}$ must act on the entanglement cut. On the other hand, as a local operator, each application of $H_\text{EF}$ can only move/manipulate the entanglement cut locally. To connect $\ket{i}$ and $\ket{j}$ states, whose entanglement cuts are separated by at least the distance of $x$, the most efficient strategy is to ``ride on the cut'', i.e.~the successive application of $H_\text{EF}$ will have to keep pushing the entanglement cut from $j$ to $i$ and always acts on the ``front cut'', so as to consume the least number of steps and to make the leading order contribution in the OTOC.

As the $F$ operator sprinkles spin-flips to the initial state $\ket{j}$, there will be multiple entanglement cuts in the resulting state $F\ket{j}$ in general. The subsequent action of $H_\text{EF}^k$ will have to clear up these entanglement cuts and bring the state to $\ket{i}$. Since $H_\text{EF}$ is a sum of local operators, it can only move/manipulate the entanglement cut locally. All the allowed processes are listed in \figref{fig: dynamics}. Crucially, the pair annihilation process is prohibited, meaning that the only way to reduce the number of entanglement cuts is the triple fusion process, which require us to first bring a pair of entanglement cuts close to the third one. Therefore the most efficient way (taking the least power of $H_\text{EF}$) to take $F\ket{j}$ to $\ket{i}$ is to sweep the right-most entanglement cut from site-$j$ to site-$i$ (assuming $j>i$). For example, as illustrated in \figref{fig: moves}(b), suppose $F$ does not act on $\ket{j}$ (which happens with weight 1), the subsequent $H_\text{EF}$ has to move the pair of entanglement cuts from site-$j$ to site-$i$ step by step, which amounts to a sequence of $w$-moves. Any additional entanglement regions between sites $i$ and $j$ can be eliminated by the sweeping protocol, as illustrated in \figref{fig: moves}(c). The rule is that when the right-most entanglement cut is adjacent to another cut, they move together to the left as a pair via the pair hopping process ($w$-move). Otherwise, the right-most entanglement cut will move to the left by itself ($v$-move). The cut hopping process will shrink the current entanglement region bounded by the moving entanglement cut, until the right entanglement cut meets its left partner and becomes a pair again. However, if $F$ generates additional entanglement regions outside the range of $i$ to $j$, one will have to take additional steps to eliminate those entanglement regions, which will only contributes to higher order expansions of the OTOC. So we will not consider those cases for now, as we are interested in the leading order contribution. 

Given the above protocol, we can see from \figref{fig: moves}(c), each $v$-move corresponds to a spin-flip introduced by $F$, so the $(-v)$ amplitude will always accompanied by a factor $d$ (coming from flipping a spin with operator $F$). The remaining steps will be implemented by $w$-move with weight $w$. If we sum over all possibilities, it seems that the inner product $\bra{i} H_\text{EF}^x F\ket{j}$ should be given by
\eq{\bra{i} H_\text{EF}^x F\ket{j}\stackrel{?}{=}(w-v d)^x.}
This answer is almost correct except for a small caveat at the initial step. The initial move of the right-most entanglement cut can be caused by both $H_\text{EF}$ and $F$. \figref{fig: moves}(d,e) show the $H_\text{EF}$ driven initial moves (where $F$ does nothing to the spin at site-$j$). \figref{fig: moves}(d,e) show the $F$ driven initial move, where a spin-flip is acted on site-$j$ to move the right-most entanglement cut. For subsequence moves, \figref{fig: moves}(d,e) will not be available, because $F$ operator can only be applied once at the very beginning. Due to the additional contribution from \figref{fig: moves}(d,e) in the initial step, the weight associate to the initial step is modified from $(w-vd)$ to $(w-2vd+ud^2)$. Therefore the correct answer is given by
\eq{\bra{i} H_\text{EF}^x F\ket{j}=(w-2vd+ud^2)(w-v d)^{x-1}.}
According to setting of the parameters $u,v,w$ in \eqnref{eq: uvw2}, we have $w-vd=-g\cosh\beta$ and $w-2vd+ud^2=-g\cosh\beta (1-d^2)$, hence
\eq{\label{eq: HFx}\bra{i} H_\text{EF}^x F\ket{j}=-(d^2-1)(-g\cosh\beta)^x.}

We can further evaluate the next order term $\bra{i} H_\text{EF}^{x+1} F\ket{j}$ following the similar strategy. The result is
\eq{\label{eq: HFx+1}\bra{i} H_\text{EF}^{x+1} F\ket{j}=(x d^2(1+\tanh^2\beta)+2(d^2-x-1)\tanh\beta)(-g\cosh\beta)^{x+1}.}
Substitute the results \eqnref{eq: HF0}, \eqnref{eq: HFx}, \eqnref{eq: HFx+1} into \eqnref{eq: OTOC1}, we obtain the OTOC to the leading orders in time
\eq{\OTOC(x,t)=1-(1-d^{-2})\frac{(tg\cosh\beta)^x}{x!}+(x (1+\tanh^2\beta)+2(1-(x+1)d^{-2})\tanh\beta)\frac{(tg\cosh\beta)^{x+1}}{(x+1)!}+\scO(t^{x+2}).}

\end{document}